\documentclass[preprint,onecolumn,nofootinbib]{revtex4}
\usepackage{graphicx}
\pdfoutput=1
\usepackage{epsfig}
\usepackage{amsmath}
\usepackage{amsfonts}
\usepackage{amssymb}
\usepackage{color}
\usepackage{dcolumn}
\usepackage{hyperref}
\usepackage{multirow}
\providecommand{\U}[1]{\protect\rule{.1in}{.1in}}

\newcommand{\f}{\begin{equation}}
\newcommand{\ff}{\end{equation}}
\newcommand{\fa}{\begin{eqnarray}}
\newcommand{\ffa}{\end{eqnarray}}
\begin{document}
\title{Holographic Shear Viscosity in Hyperscaling Violating Theories without Translational Invariance}
\author{Yi Ling $^{1,2}$}
\email{lingy@ihep.ac.cn}
\author{Zhuoyu Xian $^{1}$}
\email{xianzy@ihep.ac.cn}
\author{Zhenhua Zhou $^{1}$}
\email{zhouzh@ihep.ac.cn}
\affiliation{$^1$ Institute of High Energy Physics, Chinese Academy of Sciences, Beijing 100049, China\ \\
$^2$ Shanghai Key Laboratory of High Temperature Superconductors,
Shanghai, 200444, China}

\begin{abstract}
In this paper we investigate the ratio of shear viscosity to
entropy density, $\eta/s$, in hyperscaling violating geometry with
lattice structure. We show that the scaling relation with
hyperscaling violation gives a strong constraint to the mass of
graviton and usually leads to a power law of temperature,
$\eta/s\sim T^\kappa$. We find the exponent $\kappa$
can be greater than two such that the new bound for viscosity
raised in \cite{Hartnoll:2016tri} is violated. Our above
observation is testified by constructing specific solutions with
UV completion in various holographic models. Finally, we compare the boundedness of $\kappa$ with the behavior of entanglement entropy and conjecture a relation between them.
\end{abstract}
\maketitle

\section{Introduction}
\subsection{Motivation}
In holographic approach the Kovtun-Son-Starinets (KSS) bound for
the ratio of shear viscosity to entropy density is formulated
as \cite{Kovtun:2004de}
\begin{equation}\label{constantbound}
\frac{\eta}{s}\geq \frac1{4\pi}.
\end{equation}
Examples violating KSS bound have been proposed in the context of
holographic models with anisotropy, for instance
in \cite{Rebhan:2011vd,Ge:2014aza,Mateos:2011ix,Mateos:2011tv}, where a lower bound can be
found for the longitudinal shear viscosity in a strongly coupled
anisotropic plasma.

Recently, it is found in
\cite{Hartnoll:2016tri,Burikham:2016roo,Liu:2016njg,Alberte:2016xja,Davison:2014lua}
that this ratio is also violated when the translational invariance
is \textit{isotropically} broken in holographic theories with
lattices, massive gravity or magnetic charges,
although in this circumstance the shear viscosity does not have a
hydrodynamical interpretation and is defined by Kubo Formula (\ref{viscocitydef}), but quantifies the rate of entropy
production \cite{Hartnoll:2016tri}. A key observation in this
direction is that the introduction of lattices is equivalent to
give mass to graviton \cite{Hartnoll:2012rj,Blake:2013owa}, such
that the fluctuations of metric components become massive, giving
rise to a lower value for the viscosity bound at finite
temperature. Especially, when the lattice effect is not vanishing
in the far IR, the ratio of viscosity to entropy density
approaches to zero with a power law of temperature at leading
order
\begin{equation}\label{etaspowerlaw}
\frac{\eta}{s}\sim T^\kappa, \quad \text{as}\quad T\to0,
\end{equation}
with $0<\kappa\leq2$, where the upper bound for $\kappa$ being 2
comes from a suggested bound for the entropy production over
`Planckian time'. In our current paper we will further disclose
that this power law of $\eta/s$ is the reflection of scaling relation which emerges in the far IR.

It is very intriguing to testify whether the shear viscosity bound
proposed in \cite{Hartnoll:2016tri} holds in generic
circumstances. Motivated by this, we intend to investigate this
issue in holographic models whose background is the hyperscaling
violating geometry. In the past few years, non-relativistic
holography has extensively been studied in literature
\cite{Kachru:2008yh,Taylor:2008tg,Gubser:2009cg}, among of which
gravitational geometry enjoys the symmetry of Lifshitz fixed point
and is called Lifshitz geometry. Its time coordinate scales as
the power of space coordinate with order $z$, where $z$ is
the dynamical critical exponent. The scaling behavior has been
found in some quantum critical phenomena \cite{Sachdev:1999QPT}.
Later, a more general scaling metric conformal to the Lifshitz
one, has been realized in effective
Einstein-Maxwell-Dilaton(EMD) theories \cite{Gubser:2009qt,
Cadoni:2009xm, Goldstein:2009cv, Charmousis:2010zz,
Gouteraux:2012yr, Perlmutter:2010qu, Gouteraux:2011ce,
Iizuka:2011hg,Ogawa:2011bz, Huijse:2011ef, Alishahiha:2012cm,
Bhattacharya:2012zu,Kiritsis:2015oxa,Gouteraux:2011qh}.
Hyperscaling violation presents in those theories, since
both actions and metrics are rescaled following to a
rescaling of space, characterized by a hyperscaling violation
exponent $\theta$. In the perspective of thermodynamics, a system
with hyperscaling violation in $d$-dimensional space behaves like
the system living in a space with an effective spatial dimension
$d_\text{eff}=d-\theta$ \cite{Dong:2012se}.

Furthermore, when adding isotropic axions to the EMD model, one finds that translational invariance is broken while hyperscaling violation still holds \cite{Donos:2014uba,Gouteraux:2014hca}. A finite DC conductivity at finite temperature is obtained. A power-law behavior of conductivity with respect to low frequency and low temperature is also found, which is controlled by the scaling relation in the IR.

In this paper we intend to investigate the scaling behavior of the shear
viscosity in EMD-Axion models with hyperscaling
violation. We will concentrate on the scaling relation of IR
geometry at low temperature and then demonstrate that this
relation controls the temperature behavior of $\eta/s$.
Remarkably, we find that in a large class of holographic
models with hyperscaling violation, the exponent $\kappa$
can be greater than 2 such that the new bound proposed in
\cite{Hartnoll:2016tri} for the viscosity is violated. To make our
paper logically clear and concise, we would like to organize the
paper as follows, with a brief summary on the results of each
section.

\subsection{Summary}
In Section \ref{Sectionscalingofeta}, the
scaling behavior of $\eta/s$ is studied in a generic holographic framework
with hyperscaling violation. We prove that it is determined by a
nontrivial scaling dimension of spatial parts of energy-momentum
tensor operator $\hat{T}^{xy}$ in boundary theory when the breaking of translational invariance is relevant in the IR. This scaling
dimension is determined by the mass of graviton.

In Section \ref{SectionHVlatticescale}, we focus on EMD-Axion
theory with isotropic and relevant axion and derive the exponent
 $\kappa$ in (\ref{etaspowerlaw}). It turns out that
 $\kappa$ can be expressed as a function of spatial dimension
$d$ of the boundary theory, dynamical critical exponent $z$,
hyperscaling violating exponent $\theta$ and a positive number
$e^2$, which is defined as the ratio of Maxwell term and one of
the lattice terms in the Lagrangian. Specifically, we have
\begin{equation}\label{scaleformula}
\frac{\eta}{s}\sim T^{\frac{d+z-\theta}{z} \left(-1+\sqrt{\frac{8
(z-1)}{(d+z-\theta) (1+e^2)}+1}\right)},
\end{equation}
where parameters $(z,\theta)$ are subject to the constraints in
hyperscaling violating theory such as the null energy condition.
The above formula can reproduce the results presented in
\cite{Hartnoll:2016tri} when $\theta=0$. Novel phenomena emerge
when $\theta\neq 0$. Firstly, the exponent $\kappa$ here can be
greater than 2, violating the new bound (\ref{etaspowerlaw})
raised in \cite{Hartnoll:2016tri}. Secondly, $\kappa$ can be
negative. When $\kappa<0$, it describes the power law of the
viscosity in high temperature limit.

In Section \ref{Sectionlatticesgeometry} and
\ref{Sectionetageometry}, we numerically construct specific
background solutions which interpolate between $AdS_4$ in the UV and
hyperscaling violating geometry in the IR in Einstein-Dilaton-Axion (ED-Axion) model. Our
numerical results for the exponent $\kappa$ agree with the analytical
formula (\ref{scaleformula}).

In Section \ref{Sectiondiscussion}, we discuss the relation
between the bound of $\eta/s$ and the behavior of entanglement
entropy in hyperscaling violating theories, which may shed
light on understanding the underlying reasons leading to the
violation of the viscosity bound. Finally, we give some open
questions for further investigation.

\section{Scaling behavior of viscosity in hyperscaling violating geometry}\label{Sectionscalingofeta}
We adopt the following definition of shear viscosity in an
isotropic system.
\begin{equation}\label{viscocitydef}
\eta=\lim_{\omega\to 0}\frac1{\omega}\text{Im} G^R_{\hat{T}^{xy}\hat{T}^{xy}} (\omega,k=0),
\end{equation}
where $x,y$ are any two different spatial coordinates ($d\geq2$) and
$\hat{T}^{xy}$ is the corresponding spatial component of energy
momentum tensor. As we mentioned before, although the
hydrodynamical interpretation of this quantity is absent since the
translational invariance is broken, the definition
(\ref{viscocitydef}) is still valid and may be understood as the
quantity of entropy production.

For simplicity, we assume that the background metric and energy-momentum tensor are homogenous and isotropic in spatial directions. Thus they can be diagonalized as
\begin{equation}\begin{split}\label{metricandemtensor}
&
ds^2=-g_{tt}(r)dt^2+g_{rr}(r)dr^2+g_{xx}(r)\sum_{i=1}^d dx_i^2,
\\&
T_{\mu\nu}=\text{diag}\left(T_{tt}(r),T_{rr}(r),T_{xx}(r),\cdots,T_{xx}(r)\right).
\end{split}\end{equation}
However, we do not assume that matter fields are homogeneous. Translational invariance is broken by introducing some inhomogeneous matter fields.

The background fields satisfy the Einstein equations
\begin{equation}\label{EinsteinEquation}
R_{\mu \nu }+ \frac1{d}{g_{\mu \nu }T}-T_{\mu \nu }=0,
\end{equation}
where $T=g^{\mu\nu}T_{\mu\nu}$. As explained in
\cite{Hartnoll:2016tri}, we consider perturbation with the form as
$(\delta g)^x{}_y=h(r)e^{-i\omega t}$, whose coefficients of
boundary expansion give the Green function
$G^R_{\hat{T}^{xy}\hat{T}^{xy}}$ in the boundary theory. The
perturbation of the $(x,y)$ component of the Einstein equations
gives the shear perturbation equation
\begin{equation}\label{perturh}
\frac1{\sqrt{-g}}\partial_r(\sqrt{-g}g^{rr}\partial_r h(r)) + (g^{tt}\omega^2- m(r)^2) h(r) =0,
\end{equation}
with a square of varying mass
\begin{equation}\label{massdef}
m(r)^2= 2( g^{xx}T_{xx}-\frac{\delta T_{xy}}{\delta g_{xy}}).
\end{equation}

In standard holographic theories, there usually exists a nontrivial fixed point in the UV, which controls the high energy dynamics. Throughout this paper, we require the UV fixed point to be conformal, which is dual to AdS.

Here we are interested in shear viscosity, defined by (\ref{viscocitydef}), which is controlled by the low energy dynamics of a theory. In the holographic perspective, the scaling behavior of viscosity depends on the IR data. Here we adopt the logic of \cite{Charmousis:2010zz}. In this section, we focus on the IR geometry with hyperscaling violation, and then study the scaling behavior of viscosity. We will come back to the issue of UV completion in Subsection \ref{SubSectionUVCompletion}.

\subsection{Hyperscaling violating metrics}\label{SubSectionHVmetirc}
We consider a non-relativistic but isotropic boundary theory in
$d+1$ dimensions, which is dual to a bulk geometry with
hyperscaling violation in $d+2$ dimension. The hyperscaling
violating metric for the bulk can be written as
\begin{equation}\label{HVmetric}
ds^2= L^2 r^\frac{2\theta}{d} \left( -\frac{dt^2}{r^{2z}}+
\frac{dr^2+\sum_{i=1}^d dx_i^2}{r^2} \right),
\end{equation}
where $z$ is the dynamical critical exponent, while $\theta$ is
the hyperscaling violating exponent. $L$ is the radius of
hyperscaling violating geometry and we demand that $L^2>0$. Under
the scaling transformation $x\to\lambda x,\,r\to\lambda
r,\,t\to\lambda^z t\,$, the metric behaves as
$ds\to\lambda^{\theta/d}ds$. We may simply denote this relation as
$x\sim r\sim t^{1/z}\sim (ds)^{d/\theta}$.

Firstly, we remark that the following considerations put
constraints on the possible values of $(d,z,\theta)$ in this
hyperscaling violating metric.
\begin{enumerate}
  \item To have a well-defined IR in the bulk, we require $(\theta-d)(\theta-dz)>0$, or $\theta=d$ while $\theta\neq dz$. The condition of $\theta=d$ leads to a trivial $R^d$ in spatial directions\footnote{In \cite{Zaanen:2015oix}, it is argued that the IR geometry of extremal Reissner-Nordstr\"om(RN) black hole, $AdS_2\times R^d$, is reached by keeping $\theta=d$ and sending $z\to\infty$ in metric (\ref{HVmetric}), since entanglement entropy shows volume law when $\theta=d$ \cite{Dong:2012se,Huijse:2011ef}. However, when we only care about geometry, $AdS_2\times R^d$ can be reached by keeping $\theta$ finite and sending $z\to\infty$, such as \cite{Gouteraux:2012yr}.}.

  \item The location of IR in $r$ direction is determined by the condition that the induced line element vanishes, which leads to $r\xrightarrow{IR} 0,(\theta\geq d,\, \theta>dz)$ or $r\xrightarrow{IR}+\infty,(\theta\leq d,\, \theta<dz)$.

  \item We expect that small perturbations with modes of $\delta_0=d+z-\theta$ will generate a flow to create a small black hole with finite temperature, whose metric has the form as
      \begin{equation}\label{ThermalHVmetric}
      ds^2= L^2 r^\frac{2\theta}{d} \left( -\frac{f(r)dt^2}{r^{2z}}+\frac{dr^2}{r^2f(r)}+\frac{\sum_{i=1}^d dx_i^2}{r^2} \right), \, f(r)=1-\left(\frac{r}{r_+}\right)^{\delta_0}, \, \delta_0=d+z-\theta.
      \end{equation}
      It demands that the mode must be relevant, leading to $(d+z-\theta<0)$ if $r\xrightarrow{IR} 0$, or $(d+z-\theta>0)$ if $r\xrightarrow{IR}+\infty$. It is indeed the case in hyperscaling violation \cite{Perlmutter:2010qu,Charmousis:2010zz,Gouteraux:2012yr,Donos:2014uba,Gouteraux:2011ce,Gouteraux:2014hca,Dong:2012se}. The Hawking temperature and black hole entropy density
      \begin{equation}\label{TsHV}
      T=\frac{|\delta_0|}{4\pi}r_+^{-z}, \quad s=4\pi r_+^{\theta-d}=4\pi\left(\frac{4\pi T}{|\delta_0|}\right)^{\frac{\delta_0}{z}-1}.
      \end{equation}
      is identified with the temperature and the entropy density of the dual boundary theory. It is worthwhile to point out that both temperature and frequency scale as the inverse of time, namely $T\sim \omega \sim t^{-1}$.

  \item It is necessary to impose the Null Energy Condition (NEC), which gives rise to $(d-\theta)(d(z-1)-\theta)\geq0$ and $(z-1)(d+z-\theta)\geq0$ \cite{Dong:2012se}.
\end{enumerate}

As a result, we conclude that throughout this
paper we will only consider the system subject to the following
constraints.
\begin{equation}\begin{split}\label{constraint}
 & r\xrightarrow{IR} 0\quad (d \leq\theta \leq d+1\land d+z<\theta )\lor
(\theta >d+1\land z\leq 1),
\\
& r\xrightarrow{IR} +\infty \quad (\theta \leq 0\land z\geq 1)\lor
\left(0<\theta \leq d\land z\geq \frac{\theta }{d}+1\right).
\end{split}\end{equation}

When the black hole becomes extremal, so called extremal limit, there are two cases for the limit of temperature \footnote{The word of ``extremal" here refers
to that the black hole solution (\ref{ThermalHVmetric}) retracts its horizon $r_+$ back to the IR and
returns to the original hyperscaling  violating metric
(\ref{HVmetric}), which is equivalent to the cases in
\cite{Kiritsis:2015oxa}. When $z=0$, the limit of temperature is
subtle, so we do not discuss this case here.}.
\begin{itemize}
  \item Low temperature limit: $(d+z-\theta)z>0$. For $d+z-\theta<0, z<0$, we have $r\xrightarrow{IR} 0$ and $T\propto r_+^{-z}\to0$; while for $d+z-\theta>0, z>0$, we have $r\xrightarrow{IR} +\infty$ and $T\propto r_+^{-z}\to0$. For both cases we have $T\to 0$.
  \item High temperature limit: $(d+z-\theta)z<0$. Constraints (\ref{constraint}) give $d+z-\theta<0,z>0$, we have $r\xrightarrow{IR} 0$ and $T\propto r_+^{-z}\to\infty$.
\end{itemize}

From $s\sim T^{\frac{d-\theta}{z}}$ in hyperscaling
violating metric, we know that if the extremal limit is at
$T\to\infty$, the small black hole has negative specific
heat and is thermodynamically unstable \cite{Dong:2012se}.

In addition, investigations on the behaviors of entanglement entropy suggest that the gravitational background with $\theta>d$ might be unstable \cite{Dong:2012se}, which gives constraint stronger than (\ref{constraint}). In our paper we will ignore
it first and then come back to this issue in Section
\ref{Sectiondiscussion}.

\subsection{Scaling behavior of viscosity}\label{SubSectionScalingAnalysis}

As assumed above, the hyperscaling violating metric
(\ref{ThermalHVmetric}) is the IR limit of the background metric
in (\ref{metricandemtensor}). In the IR region, the Einstein
equations (\ref{EinsteinEquation}) give a scale relation as
$T^x{}_x\sim R^x{}_x\sim r^{-2\theta/d}$. If the breaking of translational
symmetry is (marginally) relevant in the far IR, we have
$m(r)^2\sim\nabla^2\sim g^{tt}\omega^2\sim r^{-2\theta/d}$ in
(\ref{perturh}). Similar scaling of graviton mass can be found in
\cite{Donos:2014uba,Gouteraux:2014hca,Donos:2014oha,Edalati:2012tc}.
It means that the breaking of translational invariance gives a
mass of $m(r)$ to graviton but does not break the scaling relation
above, which constrains the behavior of the mass strongly.
Furthermore, the scaling relation of hyperscaling violation is
preserved for the perturbation modes. If the breaking of translational invariance is irrelevant, $m(r)^2$ becomes subleading comparing to $\nabla^2$ in the IR. In
other words, $m(r)^2=0$ at the leading order.
While at subleading order, the irrelevant effect disturbs the
$r$ dependence of $m(r)^2$ with the involvement of other scales, which goes beyond the
following scaling analysis in the main text. For completeness, we
give a perturbation analysis and numerical calculation on
EMD-Axion model with irrelevant axion in Appendix
\ref{Appendixirr}. In the remainder of our main text, we only consider the leading order effect. At zero frequency $\omega=0$, we find the
following asymptotic expansion of $h_0(r)$
\begin{equation}\label{boundarybehavior}
h_0(r)=h_-r^{\delta_-}+\cdots+h_+r^{\delta_+}+\cdots
\end{equation}
where $\delta_-,\delta_+$ are two roots of the equation
\begin{equation}\label{scalingdim}
\delta(\delta-\delta_0)=M^2 L^2,\quad \delta_\pm=\frac12\left(\delta_0\pm\sqrt{\delta_0^2+4M^2L^2} \right),
\end{equation}
with $M^2=r^{2\theta/d}m(r)^2$ being the scaleless mass square.
The explicit form of (\ref{boundarybehavior}) is derived in
Appendix \ref{Appendix}. Eq.(\ref{scalingdim}) gives the relation
between the scaling dimension and graviton mass in the presence of
hyperscaling violation. We remark that $M^2$ should be nonnegative
($M^2\geq0$) to guarantee the stability of RG flow. Then
one of the two branches in (\ref{boundarybehavior}) is
normalizable while another is non-normalizable. For IR region, the
scaling dimension $\delta_{\hat{T}}$ of the operator
$\hat{T}^{xy}$ in dual theory should be identified with either
$\delta_-(r\xrightarrow{IR} 0)$ or $\delta_+(r\xrightarrow{IR}
+\infty)$. Taking the constraints in (\ref{constraint}) into
account, we can write $\delta_{\hat{T}}$ in an explicit form,
\begin{equation}\label{deltaT}
\delta_{\hat{T}}=\frac{\delta_0}{2}\left(1+\sqrt{1+\left(\frac{2ML}{\delta_0}\right)^2} \right),
\end{equation}
wherever the IR is located at.

Next we consider the perturbation of $h$ with frequency
$\omega$. We will find the asymptotic expansion behaves as
\begin{equation}\label{boundarybehavioromega}
h(r)= c\left( r^{\delta_0-\delta_{\hat{T}}}+\cdots+
b\mathcal{G}^R_{\hat{T}^{xy}\hat{T}^{xy}}(\omega,T)r^{\delta_{\hat{T}}}+\cdots\right),
\end{equation}
where constant $b$ plays no role in the study of scaling of Green function. Closely following the analysis presented in \cite{Dong:2012se}\footnote{The difference in our case is that the square of mass here is not a constant any more, but a quantity
scaling like the operator $\nabla^2$. This difference allows us to
define a scaleless mass.}, the
corresponding retarded Green function with $k=0$ scales as
$\mathcal{G}^R_{\hat{T}^{xy}\hat{T}^{xy}}(\omega,k=0)\sim\omega^\frac{2\delta_{\hat{T}}-\delta_0}{z}$, whose scaling dimension is $2\delta_{\hat{T}}-\delta_0$.
A general UV-IR matching procedure has been presented in \cite{Donos:2012ra,Faulkner:2009wj}, which links the imaginary part of the UV and IR Green functions as $\text{Im}G^R_{\hat{T}^{xy}\hat{T}^{xy}}(\omega,T)\propto\text{Im}\mathcal{G}^R_{\hat{T}^{xy}\hat{T}^{xy}}(\omega,T)$ at low frequency when the black hole is near the extremal limit\footnote{When the extremal limit is at $T\to\infty$, namely $\delta_0z<0$, the UV-IR matching between imaginary part of Green functions is still valid, since $\text{Im}\mathcal{G}^R_{\hat{T}^{xy}\hat{T}^{xy}}(\omega,T)$ is small at high temperature for $\delta_0z<0$, as can be seen from the exponent of $T$ in (\ref{viscosityHV}). In usual UV-IR matching, the constant $b$ is set to be 1 \cite{Faulkner:2011tm}.}.
Applying this relation, we have $\text{Im}G^R_{\hat{T}^{xy}\hat{T}^{xy}}(\omega,k=0)\sim\omega^\frac{2\delta_{\hat{T}}-\delta_0}{z}$. Then, by definition (\ref{viscocitydef}), shear viscosity scales as
$\eta\sim\omega^{\frac{2\delta_{\hat{T}}-\delta_0}{z}-1}\sim
T^{\frac{2\delta_{\hat{T}}-\delta_0}{z}-1}$, whose scaling dimension is $2\delta_{\hat{T}}-\delta_0-z$. Remind that the
entropy density scales as $s\sim T^{\frac{d-\theta}{z}}$, thus we
obtain the ratio of shear viscosity and entropy density which
scales as
\begin{equation}\label{viscosityscaling}
\frac{\eta}{s}\sim
T^\kappa=T^{\frac{2(\delta_{\hat{T}}-\delta_0)}{z}}=T^{\frac{d_\text{eff}+z}{z}\left(-1+\sqrt{1+\left(\frac{2ML}{d_\text{eff}+z}\right)^2}
\right)},
\end{equation}
where effective spatial dimension $d_\text{eff}=d-\theta$. A more detailed derivation is given in Appendix \ref{Appendix}.

For $M^2=0$, we have $\delta_{\hat{T}}=\delta_0$, $\eta/s\sim
1$, thus obtain the usual scaling dimension of $\hat{T}^{xy}$ \cite{Chemissany:2014xpa,Taylor:2015glc} and a constant $\eta/s$ bound \cite{Kovtun:2004de,Kolekar:2016pnr,Kuang:2015mlf}. While for
$M^2>0$, we have a nonzero exponent $\kappa$ and $\eta/s$
exhibits a power law of temperature. The value of
$M^2$ is model-dependent. In the presence of hyperscaling
violation, we find it is completely possible to have an exponent
$\kappa$ greater than 2 or even less than 0, under the constraints
(\ref{constraint}). We will push this point forward in next
sections.

Moreover, according to the discussion on the limit of temperature
above, when $(d+z-\theta)z>0$ ($(d+z-\theta)z<0$), we have $\kappa>0$ ($\kappa<0$), then
Eq.(\ref{viscosityscaling}) describes the low (high) temperature
behavior of $\eta/s$.

In the end of this section, we remark that our results obtained in
(\ref{viscosityscaling}) is consistent with the (weaker)
horizon formula for $\eta/s$ in $d+2$ dimension \cite{Lucas:2015vna,Hartnoll:2016tri},
\begin{equation}\label{viscosityh+d+2}
\frac{\eta}{s}=\frac1{4\pi}h_0(r_+)^2,
\end{equation}
where $h_0(r)$ is required to be equal to $1$ on the
boundary. Since the IR regular branch of $h_0(r)$ behaves as
$h_0(r)\sim r^{\delta_0-\delta_{\hat{T}}}$, after perturbing to
finite temperature (\ref{ThermalHVmetric}) we have $h_0(r_+)\sim
r_+^{\delta_0-\delta_{\hat{T}}}, T\sim r_+^{-z}$ and then
reproduce the result in (\ref{viscosityscaling}).

In next section we will consider specific models in EMD-Axion
theory in which the graviton mass can be evaluated out explicitly.

\section{Hyperscaling violating solution with lattices}\label{SectionHVlatticescale}
We work on a $(d+2)$-dimensional EMD-Axion theory whose action reads
as
\begin{equation}\label{action}
\mathcal{S}=\int dt d^dx dr\sqrt{-g}( R+ \mathcal{L}_m) ,\quad
\mathcal{L}_m = -\frac{c}{2}(\partial \phi
)^2-\frac{J(\phi)}{2}\sum_{i=1}^d(\partial\chi_i)^2+V(\phi)-\frac{Z(\phi)}{4}F^2,
\end{equation}
where $i=1,2...d$ correspond to spatial directions and $c,J(\phi),Z(\phi)\geq0$. The equations of motion
can be written as the following forms
\begin{equation}\begin{split}\label{EOM}
&
R_{\mu \nu }+ \frac1{d}{g_{\mu \nu }T}-T_{\mu \nu }=0,\quad T_{\mu
\nu
}=-\frac1{\sqrt{-g}}\frac{\delta(\sqrt{-g}\mathcal{L}_m)}{\delta
g^{\mu \nu }} =\frac12 g_{\mu \nu } \mathcal{L}_m -
\frac{\delta\mathcal{L}_m}{\delta g^{\mu \nu }},
\\&
c\nabla^2\phi-\frac{J'(\phi)}{2}\sum_{i=1}^d(\partial\chi_i)^2+V'(\phi)-\frac{Z'(\phi)}{4}F^2=0,
\\&
\nabla^\nu(Z(\phi)F_{\mu\nu})=0, \quad \nabla^\mu(J(\phi)\partial_\mu\chi_i)=0.
\end{split}\end{equation}
Here we only consider the static and isotropic solutions with matter fields
\begin{equation}\label{matter}
\phi=\phi(r),\quad \chi_i=kx_i,\quad A=A_t(r)dt
\end{equation}
and metric (\ref{metricandemtensor}), where $k$ characterizes the lattices scale. The
translational invariance is broken by the axions.
Given the action above, the square of varying mass in (\ref{massdef}) is
\begin{equation}
m(r)^2=-2g^{xx}\frac{\delta\mathcal{L}_m}{\delta g^{xx}}=J(\phi)(\partial\chi_x)^2,
\end{equation}
where $x$ refers to any one of the $d$ spatial directions and $\chi_x=kx$. The first
equality comes from that the metric is diagonal and
$\mathcal{L}_m$ is linear to the spatial components of metric. Moreover, due to the
isotropy of the background, we find that the Einstein equations in (\ref{EOM}) lead to
\begin{equation}\label{m2eqgeo}
R^x{}_x-R^t{}_t=T^x{}_x-T^t{}_t=g^{tt}\frac{\delta\mathcal{L}_m}{\delta
g^{tt}}-g^{xx}\frac{\delta\mathcal{L}_m}{\delta g^{xx}}
=-\frac{Z(\phi)}{4}F^2+\frac12J(\phi)(\partial\chi_x)^2\equiv\frac{1+e(r)^2}{2}m(r)^2,
\end{equation}
where
\begin{equation}\label{chargetomassratio}
e(r)^2=g^{tt}\frac{\delta\mathcal{L}_m}{\delta g^{tt}}\left/\left(-g^{xx}\frac{\delta\mathcal{L}_m}{\delta g^{xx}}\right)\right.
=-\frac{Z(\phi)}{4}F^2\left/\left(\frac12J(\phi)(\partial\chi_x)^2\right)\right.\geq0.
\end{equation}
Note that the l.h.s. of (\ref{m2eqgeo}) is a purely geometric
quantity. When $e(r)^2=0$, namely the Maxwell term in
$\mathcal{L}_m$ vanishing, both $m(r)^2$ and $h(r)$ in
(\ref{perturh}) depend only on the bulk geometry of the background
as in appearance. It reflects a strong constraint to the mass of graviton, while the presence of the Maxwell field may modify it.

We assume that a hyperscaling violating solution to the equations of motion (\ref{EOM}) exists in the far IR
\begin{equation}\label{HV}
ds^2= L^2 r^\frac{2\theta}{d} \left( -\frac{dt^2}{r^{2z}}+
\frac{dr^2+\sum_{i=1}^d dx_i^2}{r^2} \right),\quad
A=Qr^{\zeta-z}dt,\quad e^\phi=e^{\phi_0}r^{\phi_1},
\end{equation}
where $\zeta$ is charge anomaly. This assumption is not
difficult to reach. Hyperscaling violation emerges in the IR of
many isotropic extremal solutions of the EMD-Axion Theory
(\ref{action}) above, except some (possibly) non-scaling
solutions such as insulating phase of Q-lattices
\cite{Donos:2013eha}. Especially, solutions with ($\theta\neq0$)
can be found by choosing the form of potentials $V(\phi), J(\phi)$
and $Z(\phi)$ as
\begin{equation}
V(\phi)\propto e^{\alpha\phi}, \quad J(\phi)\propto e^{\beta\phi}, \quad Z(\phi)\propto e^{\gamma\phi},
\end{equation}
when $\phi\to\pm\infty$
\cite{Gouteraux:2012yr,Donos:2014uba,Gouteraux:2014hca}. If the
potentials have some subleading exponential terms which deviate
from an exponential form of $\phi$, the solution above is valid
only at leading order \cite{Kiritsis:2015oxa}. But for our purpose
it is enough to discuss the scaling behavior of the leading terms
here. It is natural to demand $Q^2\geq0$ and $L^2>0$ which
give more requirements to a certain model.

From the scaling relation of hyperscaling violation, it is
reasonable to compare the scaling of the Maxwell term and
axion term in the Lagrangian (\ref{action}) or Eq.(\ref{m2eqgeo}). If they have the same
scaling, $e(r)^2$ reaches a finite and scaleless constant.
Otherwise, at least one of them should be subleading, then $e(r)^2\to0$ or $+\infty$ in the far IR. So we denote
$e(r)^2=e^2$. Then with the use of the metric, it is easy to
obtain
\begin{equation}
R^x{}_x=\frac{\delta_0(\theta-d)}{dL^2}r^{-2\theta/d},\quad
R^t{}_t=\frac{\delta_0(\theta-dz)}{dL^2}r^{-2\theta/d}.
\end{equation}
Substituting them into (\ref{m2eqgeo}), we obtain the square of
scaleless mass as
\begin{equation}\label{massEMDA}
M^2=\frac{2\delta_0(z-1)}{(1+e^2)L^2}.
\end{equation}
The positivity of $M^2$ is guaranteed by one of the NEC,
namely $(z-1)(d+z-\theta)\geq0$. Finally, substituting the expression
of mass into (\ref{viscosityscaling}), we have
\begin{equation}\label{bound}
\frac{\eta}{s}\sim T^{\frac{d_\text{eff}+z}{z}
\left(-1+\sqrt{1+\frac{8 (z-1)}{(d_\text{eff}+z)
(1+e^2)}}\right)}.
\end{equation}

Here we have obtained the specific form for the exponent $\kappa$
in hyperscaling violating solutions with the action (\ref{action}),
which in general is a function of effective spatial dimension
$d_{\text{eff}}$, dynamical critical exponent $z$ and a number
$e^2$, which is formally defined as the ratio of the Maxwell term
and one of the lattice terms.

As a check, here we may immediately apply our formula in (\ref{bound}) to some specific models previously discussed in
\cite{Hartnoll:2016tri}.
\begin{itemize}
  \item Neutral linear axion model. Its extremal IR geometry is neutral $AdS_2\times R^2$, corresponding to the situation of $(d=\theta=2,z\to+\infty,e^2=0)$. We get that $\frac{\eta}{s}\sim T^2$.
  \item Charged linear axion model. Its extremal IR geometry is charged $AdS_2\times R^2$, corresponding to the situation of $(d=\theta=2,z\to+\infty,e^2=\frac{4\mu^2}{\alpha^2\gamma^2})$. We get that $\frac{\eta}{s}\sim T^{-1+\sqrt{1+\frac{8}{1+4\mu^2/(\alpha^2\gamma^2)}}}$. \footnote{The action (\ref{action}) and matter fields (\ref{matter}) in the notation of \cite{Hartnoll:2016tri} are $c=0,J(\phi)=1,Z(\phi)=\frac4{\gamma^2}$ and $\phi=0,\chi_i=\alpha x_i, A_t|_\text{bdy}=\mu$.}
  \item Neutral Q-lattices. Its extremal IR geometry is neutral $AdS_4$, where dilaton vanishes exponentially and the translational invariance is recovered. It corresponds to the situation of $(d=2,\theta=0,z=1,e^2=0)$. We get that $\frac{\eta}{s}\sim 1$.
  \item Metallic phase of charged Q-lattices. Its extremal IR geometry is charged $AdS_2\times R^2$ with irrelevant lattices, corresponding to the situation of $(d=\theta=2,z\to+\infty,e^2\to+\infty)$. We get that $\frac{\eta}{s}\sim 1$.
\end{itemize}
All the results above match the low temperature behavior as
described in \cite{Hartnoll:2016tri}. There is no surprise since
their extremal IR geometries belong to the special class of
hyperscaling violating geometry with $\theta=0$, and the mass of
graviton is restricted by the scaling relation.

\begin{figure}
  \centering
  \includegraphics[width=300pt]{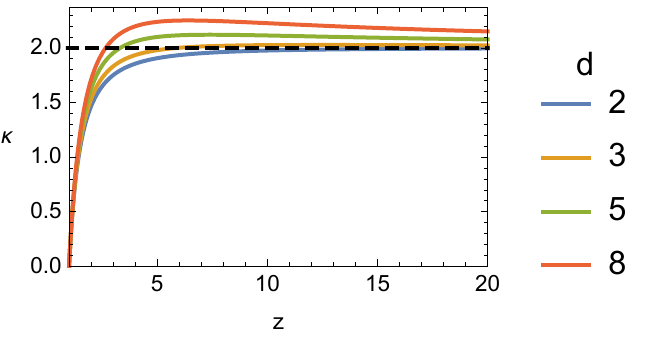}\\
  \caption{The exponent $\kappa$ of $\frac{\eta}{s}\sim T^\kappa$ as a function of $z$, with $\theta=0$ and $e^2=0$, for $d=2,3,5,8$. $\kappa$ can be greater than $2$ at finite $z$, when $d>2$. }\label{boundvszd}
\end{figure}

Definitely, we may provide more generic holographic models with
attractive features in the framework of hyperscaling violating
theory, among of which we would like to discuss several special
situations as listed below.
\begin{itemize}
\item $z=1$. Geometries are conformal to $AdS_{d+2}$, whose
lorentz symmetry is preserved but hyperscaling relation may be
violated (if $\theta\neq 0$). We obtain $\eta/s\sim 1$ as a usual
constant bound \cite{Kolekar:2016pnr}.

\item $\theta=0, e^2=0$.
Geometries are Lifshitz and the constraints (\ref{constraint})
reduce to $z\geq1$. When $d=2$, $\kappa$ is a monotonously
increasing function of $z$ and reach its maximum with $\kappa=2$
at $z\to\pm\infty$, which is consistent with the new bound
proposed in \cite{Hartnoll:2016tri}. When $d>2$, $\kappa$ is not
monotonous any more and its maximal value can be greater than $2$
at finite $z$, as shown in Figure \ref{boundvszd}, which is a
signal of violating the new bound above. As a matter of fact,
we remark that the vanishing of $e^2$ is not necessary here.
A nonzero but small $e^2$ can make $\kappa$ greater than 2 as well.
Solutions with relevant axions and the full Lifshitz symmetry have been found in \cite{Gouteraux:2014hca}. Flows from AdS to this kind of fixed points are worth building.

\item $e^2\geq0$. Under the constraints (\ref{constraint}),
we find $\kappa\frac{\partial\kappa}{\partial(e^2)}\leq0$, where
the equality holds up if and only if $\kappa=0$. It means
that in the low (high) temperature limit regions, the charge is
always reducing (enlarging) the exponent, except $\kappa=0$.

\item $z\to\infty$, while keeping $\theta$ finite. Geometry is
$AdS_2\times R^d$, and we get $\frac{\eta}{s}\sim
T^{-1+\sqrt{1+\frac{8}{1+e^2}}}$, whose exponent is not greater
than 2.

\item $z,\theta\to\infty$, while keeping $-\frac{\theta}{z}=\bar{\eta}$ fixed. Constraints
(\ref{constraint}) lead to $\bar{\eta}\geq0$. Geometry is
conformal to $AdS_2\times R^d$, so called ``$\eta$-geometry", describing the semi-local quantum criticality
\cite{Donos:2014uba,Gouteraux:2012yr,Gouteraux:2011ce,Gouteraux:2014hca}.
We get $\frac{\eta}{s}\sim T^{(1+\bar{\eta})
\left(-1+\sqrt{1+\frac{8}{\left(e^2+1\right) (1+\bar{\eta})}}\right)}$. We build model for this situation in
Section \ref{Sectionetageometry}.
\end{itemize}

\begin{figure}
  \centering
  \includegraphics[width=200pt]{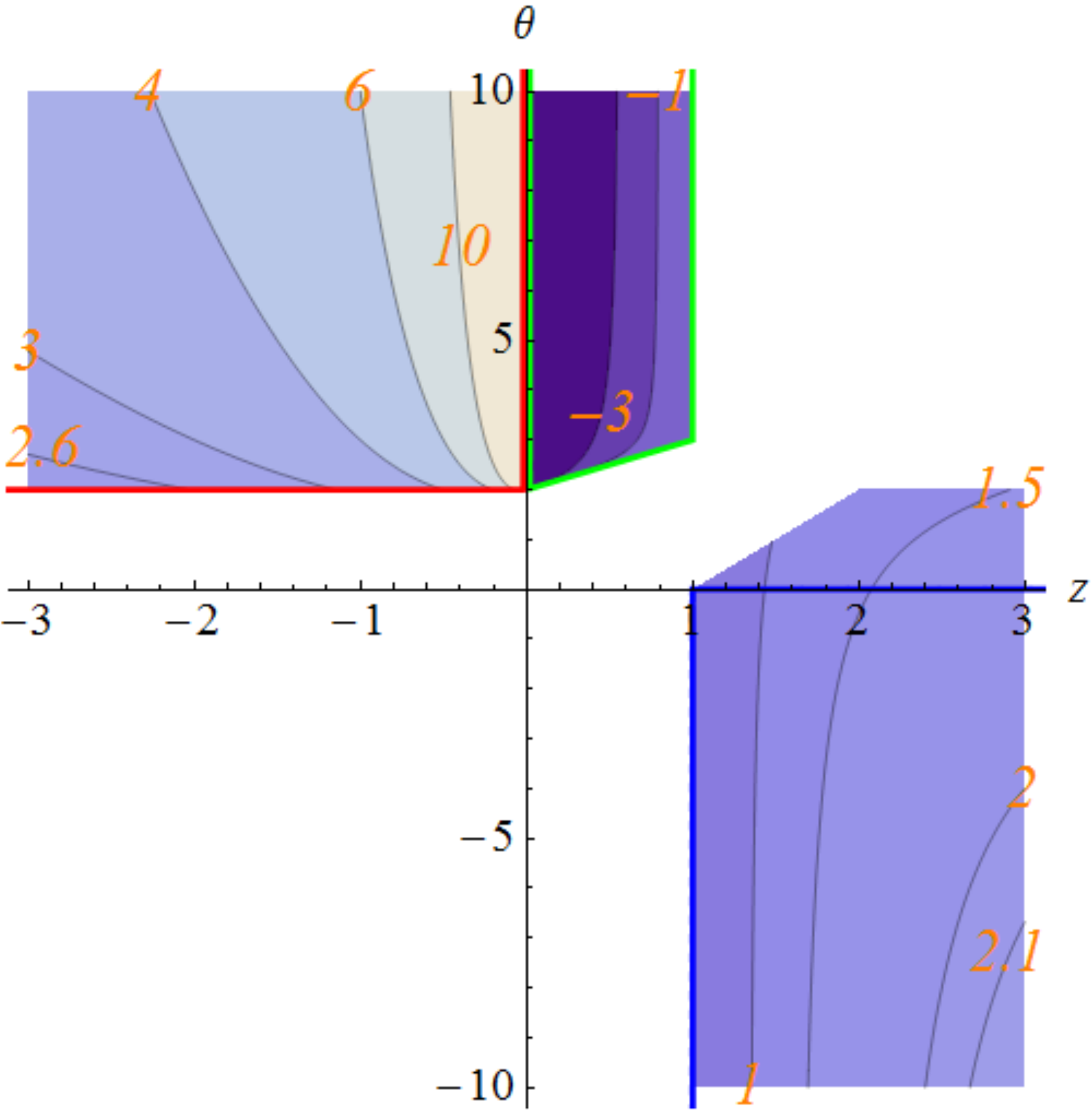}\\
  \caption{Contour plot of the exponent $\kappa$ of $\frac{\eta}{s}\sim T^\kappa$ over $(z,\theta)$ plane with $d=2$ and $e^2=0$. The left-upper region corresponds to $r\xrightarrow{IR} 0$, which is divided into Region A (surrounded by red line) and Region B (surrounded by green line) as discussed in Section \ref{Sectionlatticesgeometry}. The right-lower region corresponds to $r\xrightarrow{IR} +\infty$ and contains Region C (surrounded by blue line). The blank region violates the constraints (\ref{constraint}).}\label{contourztheta}
\end{figure}

In Figure \ref{contourztheta}, we plot the value of $\kappa$ as a
function of $(z,\theta)$ in the allowed region with $d=2$. It is
noticed that the value of $\kappa$ can be greater than $2$.

Up to now, we have only concentrated on the extremal IR geometry
with hyperscaling violation by an analytical consideration, with a
signal that the bound for $\kappa$ could be violated in some
situations. It is still questionable if we could explicitly
construct such kind of black hole solutions with UV completion at
finite temperature. As a matter of fact, we point out with caution
that model building may not be realized for all the
parameters because the stability of the IR scaling solution and the
existence of UV completion must be taken into account, as well as
other natural requirements, such as $L^2>0,Q^2\geq0$. Therefore,
in next two sections we will address this issue by numerically solving
the equations of motion and constructing explicit black hole
solutions in which the new bound proposed in
\cite{Hartnoll:2016tri} is violated.

\section{Isotropic dilaton-axion lattices with finite $(z,\theta)$}\label{Sectionlatticesgeometry}
As explained in Section \ref{SectionHVlatticescale}, charge always reduces the exponent $\kappa$ for $\kappa>0$, so usually the charge plays no role in discussing the upper bound of $\kappa$. For simplicity, we continue to study on neutral backgrounds with relevant axion, which has already captured the power laws of $s\sim T^\lambda$ and $\frac{\eta}{s}\sim T^\kappa$. We immediately have $e^2=0$.
\subsection{Scaling solution and stability}\label{SubSectionScaling}
We first consider the following 4-dimensional ED-Axion model
\begin{equation}\label{IRactionA}
\mathcal{S}=\int dt d^2x dr\sqrt{-g}\left[ R - \frac{c}{2}\left((\partial\phi)^2+ e^{2\phi}\sum_{i=1,2}(\partial\chi_i)^2\right) +n_1e^{\alpha\phi} \right],
\end{equation}
in which we have chosen the potentials as $J(\phi)=c e^{2\phi},\, V(\phi)=n_1e^{\alpha\phi}$. Scaling solutions have been found in \cite{Donos:2014uba,Gouteraux:2014hca}. We deduce them in the hyperscaling violating ansatz (\ref{HV}) here. Looking for a solution of hyperscaling violation with relevant axions, we require that each term in the Lagrange should scale in the same way, i.e. $R\sim e^{2\phi}(\partial\chi_i)^2\sim e^{\alpha\phi}$. So their exponents of $r$ satisfy equalities as
\begin{equation}
-\theta=2 \phi_1+2-\theta=\alpha\phi_1.
\end{equation}
We immediately have
\begin{equation}\label{solutionexp}
\theta=\alpha,\quad \phi_1=-1.
\end{equation}
Other parameters are deduced by solving the equations of motion. The result is
\begin{equation}\begin{split}\label{solution}
&
z=\frac{\alpha ^2-c-4}{2(\alpha-2) },\quad e^{\alpha \phi_0}L^2=\frac{((\alpha -6) (\alpha -2)+c) (-2 \alpha +c+4)}{2 (\alpha -2)^2 n_1},
\\&
e^{2 \phi_0}k^2=\frac{((\alpha -6) (\alpha -2)+c) (c-(\alpha -2) \alpha )}{2 (\alpha -2)^2 c}.
\end{split}\end{equation}
The above neutral solution is just the leading order solution with irrelevant current in \cite{Donos:2014uba,Gouteraux:2014hca}. It should give the same exponent $\kappa$ which only depends on the geometric parameters $(z,d_\text{eff})$. Besides, under the constraints (\ref{constraint}), $k^2\geq0$ and $L^2>0$ are satisfied if $n_1>0$.

By using (\ref{solutionexp}) and (\ref{solution}), the scaling behaviours (\ref{bound}) can be written in terms of $\alpha$ and $c$ as
\begin{equation}\label{EMDAscaling}
s\sim T^{\frac{2 (\alpha -2)^2}{-\alpha ^2+c+4}},\quad
\frac{\eta}{s}\sim T^{-\frac{(\alpha -8) \alpha -\sqrt{((\alpha -6) (\alpha -2)+c) \left(-7 \alpha ^2+8 \alpha +9 c+12\right)}+c+12}{-\alpha ^2+c+4}}.
\end{equation}

We now analyze the static modes by considering the
following perturbation about the hyperscaling violating solution.
\begin{equation}
ds^2= L^2 r^\theta \left( -\frac{(1+c_t r^\delta)dt^2}{r^{2z}}+
\frac{(1+c_r r^\delta)dr^2+(1+c_x r^\delta)(dx_1^2+dx_2^2)}{r^2}
\right), \quad e^\phi=e^{\phi_0}r^{-1}(1+c_\phi r^\delta).
\end{equation}
By solving linearized perturbation equation, we find two pairs of modes after getting rid of the trivial modes \footnote{The class of trivial modes comes from the redundance of the perturbation. They are proportional to $c_t=\theta-2z,c_r=\theta+ 2\delta,c_x=\theta-2,c_\phi=-1$ for any $\delta$, which correspond to the infinitesimal transformation $r\to r(1+\epsilon r^\delta)$ where $\epsilon<<1$.}. The two pairs of modes $\delta_\pm$ satisfying $\delta_-+\delta_+=\delta_0=2+z-\theta$. The first pair has $\delta^{(0)}_-=0$ and $\delta^{(0)}_+=\delta_0$ ($-c_t=c_r=r_+^{-\delta_0},c_x=c_\phi=0$), which correspond to rescaling of time and creating a small black hole (\ref{ThermalHVmetric}) respectively. The other pair has
\begin{equation}
\delta_\pm^{(1)}=\frac12\left(\delta _0\pm\frac{\sqrt{\delta _0 (-\theta +2 z-2) \left(\theta ^2+8 \theta -11 \theta  z+2 z (9 z-7)-4\right)}}{- \theta +2 z -2}\right).
\end{equation}
We point out that the relation $\delta_+>\delta_-$ is not implied for those two pairs of modes.

Since the location of the IR depends on $(z,\theta)$ or $(\alpha,c)$, we can not determine whether a mode is relevant or irrelevant from the sign of $\delta$. A plausible way is to check the sign of $\delta_-\delta_+$. If it is negative, then we always find that one of the pair of modes is irrelevant and stands for source, irrespective of the location of the IR. Here thanks to the constraints (\ref{constraint}), we have $\delta_-^{(1)}\delta_+^{(1)}=-\frac{2 (z-1) (-\theta +z+2) (2 z-\theta )}{-\theta +2 z-2}\leq0$, thus the scaling solution is RG stable.

The irrelevant mode among $\delta^{(1)}_\pm$ is adjusted to satisfy the boundary condition of $\phi$ on the UV boundary after UV completion; while the relevant mode of $\delta^{(0)}_+=\delta_0$ drives the extremal solution to a black hole with finite temperature. They are generally sufficient to construct a domain wall between AdS and hyperscaling violating geometry at finite temperature, which will be studied in the next subsection.

\subsection{UV completion and numerical results}\label{SubSectionUVCompletion}

As mentioned at the beginning of Section \ref{Sectionscalingofeta}, now we should do the UV completion to construct the bulk solution which is asymptotic to AdS. As explained in \cite{Charmousis:2010zz}, the UV completing process can be achieved by demanding that $e^{\alpha\phi}\to0$ in the UV of our previous solution and modifying the potential like $V(\phi)\to\frac{d(d+1)}{l^2}+n_1e^{\alpha\phi}$, where $l$ is the radius of AdS and is chosen to be 1 for convenience.

From solution (\ref{solutionexp}), we have $e^{\alpha\phi}\propto
r^{-\theta}$. The above UV completing process demands that
$r^{-\theta}\to0$ in the UV. For $r\xrightarrow{IR}0$, constraints
(\ref{constraint}) lead to $\theta>0$, then the requirement
$e^{\alpha\phi}\xrightarrow{UV}0$ is satisfied such that we can
find a flow from $AdS_4$. On the other hand, for
$r\xrightarrow{IR}+\infty$, we require $\theta<0$, which falls
into a region of the constraints (\ref{constraint}), as shown in
Figure \ref{contourztheta}. Notice that the UV completion process we adopt is not applicable to the region of $0\leq\theta<2$ . Nevertheless, we expect that other kind of UV completion is helpful, such as adopting a potential similar to the one in \cite{Ogawa:2011bz, Huijse:2011ef}. We expect to realize it in future.

Taking the different limits of temperature into account, we
conclude that the allowed values for  $(z,\theta)$  can be
classified into three regions, as summarized in Table
\ref{constraintregion1}, among of which Region A and C have
been mentioned in \cite{Donos:2014uba,Gouteraux:2014hca}. These
three regions have been marked in Figure \ref{contourztheta}.

\begin{table}
  \newcommand{\tabincell}[2]{\begin{tabular}{@{}#1@{}}#2\end{tabular}}
  \centering
\begin{tabular}{|c|c|c|c|c|c|}
  \hline
  Regions    & IR & Limit of $T$ & $(z,\theta)$ & $(\alpha,c)$ & $\kappa$    \\
  \hline \cline{1-6}
  Region A  & $r\xrightarrow{IR}0$          & $T\to 0$  & $z < 0\land \theta >2$    & $2<\alpha<\sqrt{4+c}$ & $\kappa>2$   \\
  \hline
  Region B  & $r\xrightarrow{IR}0$ & $T\to +\infty$ & $0<z\leq 1\land \theta >z+2$ & \fontsize{8pt}{\baselineskip}{\tabincell{c}{$(2<\alpha \leq 3\land -\alpha ^2+8 \alpha -12<c<\alpha ^2-4 )$\\$\lor (\alpha >3\land \alpha ^2-2 \alpha \leq c<\alpha ^2-4)$}} & $\kappa\leq0$ \\
  \hline
  Region C  & $r\xrightarrow{IR} +\infty$   & $T\to 0$  & $\theta < 0\land z \geq 1$   & $\alpha <0\land c\geq \alpha ^2-2 \alpha$ & $0\leq\kappa<4$   \\
  \hline
\end{tabular}
  \caption{Three regions with different locations of IR and different limits of temperature when black holes become extremal. In the last column, we show the ranges of $\kappa$ by using (\ref{bound}).}\label{constraintregion1}
\end{table}

As a result, we choose the following action with UV completion for Region A and Region B.
\begin{equation}\label{UVactionA}
\mathcal{S}=\int dt d^2x dr\sqrt{-g}\left\{ R+ 6 + \frac{4c}{\alpha^2}\sinh(\frac{\alpha}{2}\phi)^2 - \frac{c}{2}\left[(\partial\phi)^2 + 4 \sinh ^2(\phi ) \sum_{i=1,2}(\partial\chi_i)^2 \right]\right\}.
\end{equation}
The form of potential $V(\phi)$ imitates that in (6.1) in \cite{Kiritsis:2015oxa}. When $\phi\to\infty$, we have $n_1=c/\alpha^2$.

The action admits an $AdS_4$ vacuum with unit radius. Since $V(\phi)=6 + \frac{4c}{\alpha^2}\sinh(\frac{\alpha}{2}\phi)^2=6+c\phi^2+...$, the square of mass of dilaton is $-2$ on the boundary. Then we choose the conformal weight of its dual operator as $\Delta=1$.

The action does not allow a zero temperature solution with the near horizon geometry of $AdS_2\times R^2$ and $\phi=0$, since the term in front of axions, namely $\sinh ^2(\phi )$, vanishes when the dilaton vanishes\footnote{It is pointed out by \cite{Donos:2014uba} that a ground state of $AdS_2\times R^2$ with nonvanishing dilaton and axions does exist when $\alpha=2$, which corresponds to $z\to\infty$ here, as can be seen from (\ref{solution}). However, we do not study it here.}.

We adopt the following ansatz for numerical calculation.
\begin{equation}\label{ansatz}
ds^2=\frac1{u^2}\left(-(1-u)U(u)e^{-S(u)}dt^2+\frac{du^2}{(1-u)U(u)}+ dx_1^2 + dx_2^2 \right),\quad \phi=\phi(u),\quad \chi_{1,2}=kx_{1,2}.
\end{equation}
The AdS boundary is located at $u=0$. The horizon has been rescaled to $u=1$ such that the temperature and entropy density are $T=\frac1{4\pi}U(1)e^{-S(1)/2}$ and $s=4\pi$. The free energy density is $f=\epsilon-Ts$, where the energy density $\epsilon$ comes from the boundary expansion $(1-u)U(u)e^{-S(u)}=1+\cdots-\frac{\epsilon}{2}u^3+\cdots$. The dimensionless temperature, entropy density and free energy density are
\begin{equation}
\tilde{T}=\frac{T}{k},\quad \tilde{s}=\frac{s}{k^2}, \quad \tilde{f}=\frac{f}{k^3},
\end{equation}
where $k$ is the lattice number. We set the AdS boundary conditions as $U(0)=1,\,S(0)=0,\,\phi''(0)-\tau \phi'(0)=0$, while impose the regularity boundary condition  at horizon. As a result, each solution  here is  parameterized by  two quantities, $(\tau, T/k)$, where $\tau$ is a dimensionless parameter specified by the AdS boundary condition of $\phi$.

Now we numerically build up the background solution and then solve the perturbation equation of $h(u)$ (\ref{perturh}). Changing $T/k$ with a fixed $\tau$, we can numerically construct hyperscaling violating solutions in the IR only within a certain range of $\tau$. Finally we calculate $\eta/s$ numerically with the use of Eq. (\ref{viscosityh+d+2}). We verify the power law behavior of $s$ and $\eta/s$ for some values of $(\alpha,c)$ in Region A and Region B, which is independent of the value for $\tau$. We find $\frac{\eta}{s}\leq\frac1{4\pi}$ in all the cases, and the equality holds up only when the black hole reaches the limit which is opposite to the extremal limit. We give some remarks as listed below.

\begin{figure}
  \centering
  \includegraphics[width=230pt]{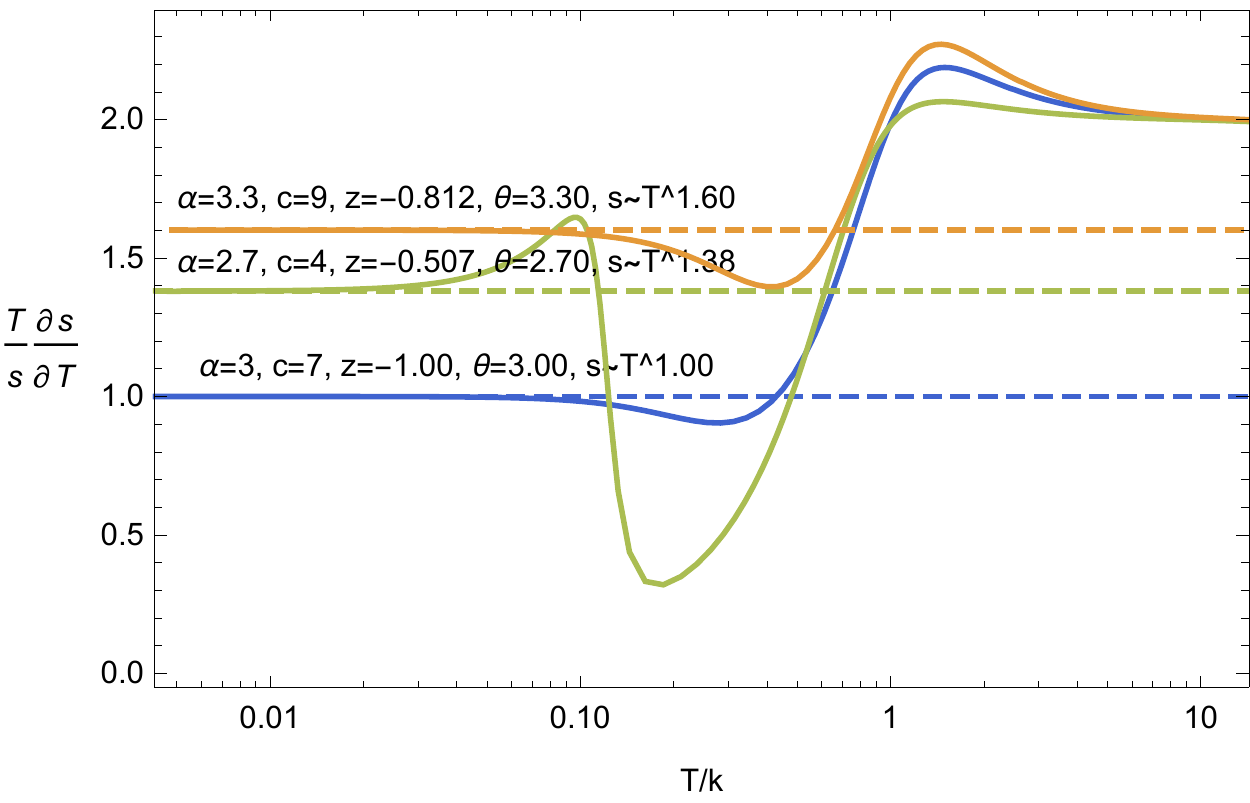}
  \includegraphics[width=230pt]{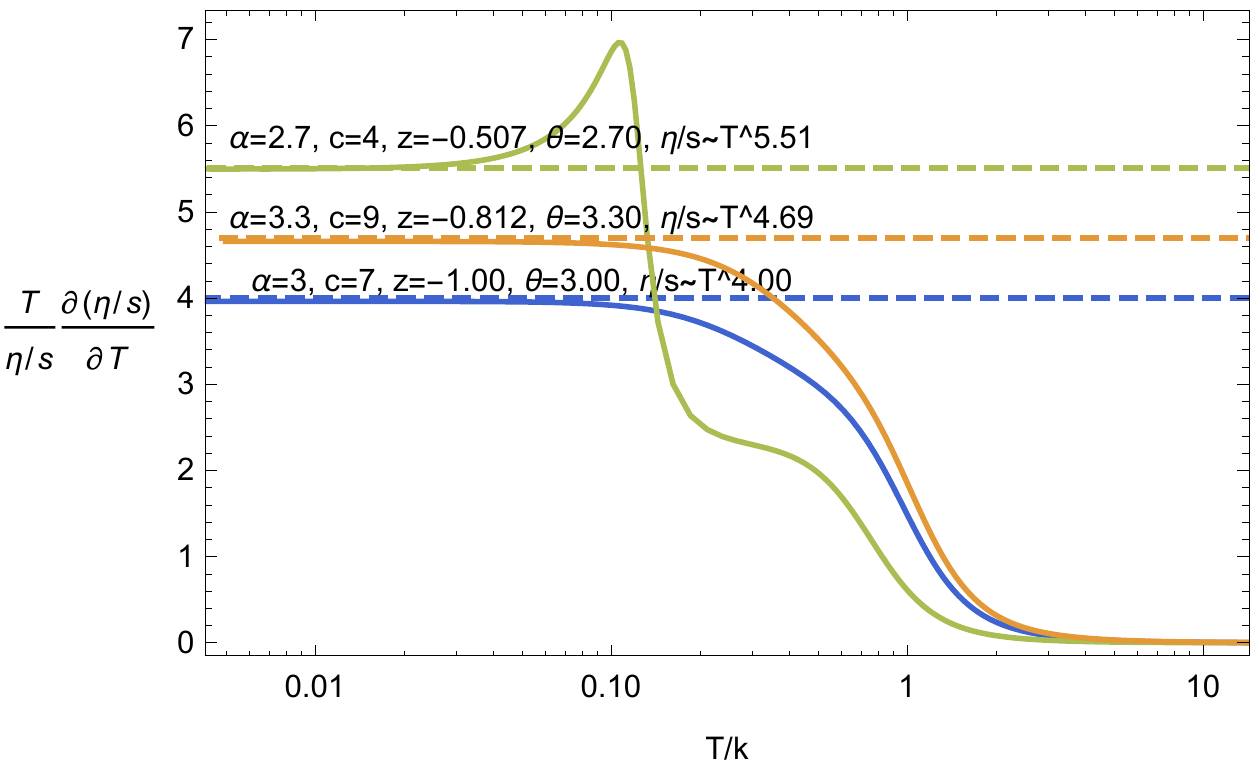}
  \caption{The scaling exponents of $s\sim T^\lambda$ (the left plot) and $\frac{\eta}{s}\sim T^\kappa$ (the right plot) as a function of $T/k$ in Region A. Solid lines represent the numerical results; dashed lines represent analytical results from (\ref{EMDAscaling}). The extremal limit is at $T\to0$.}\label{TpowersA1}
\end{figure}

\begin{figure}
  \centering
  \includegraphics[height=140pt]{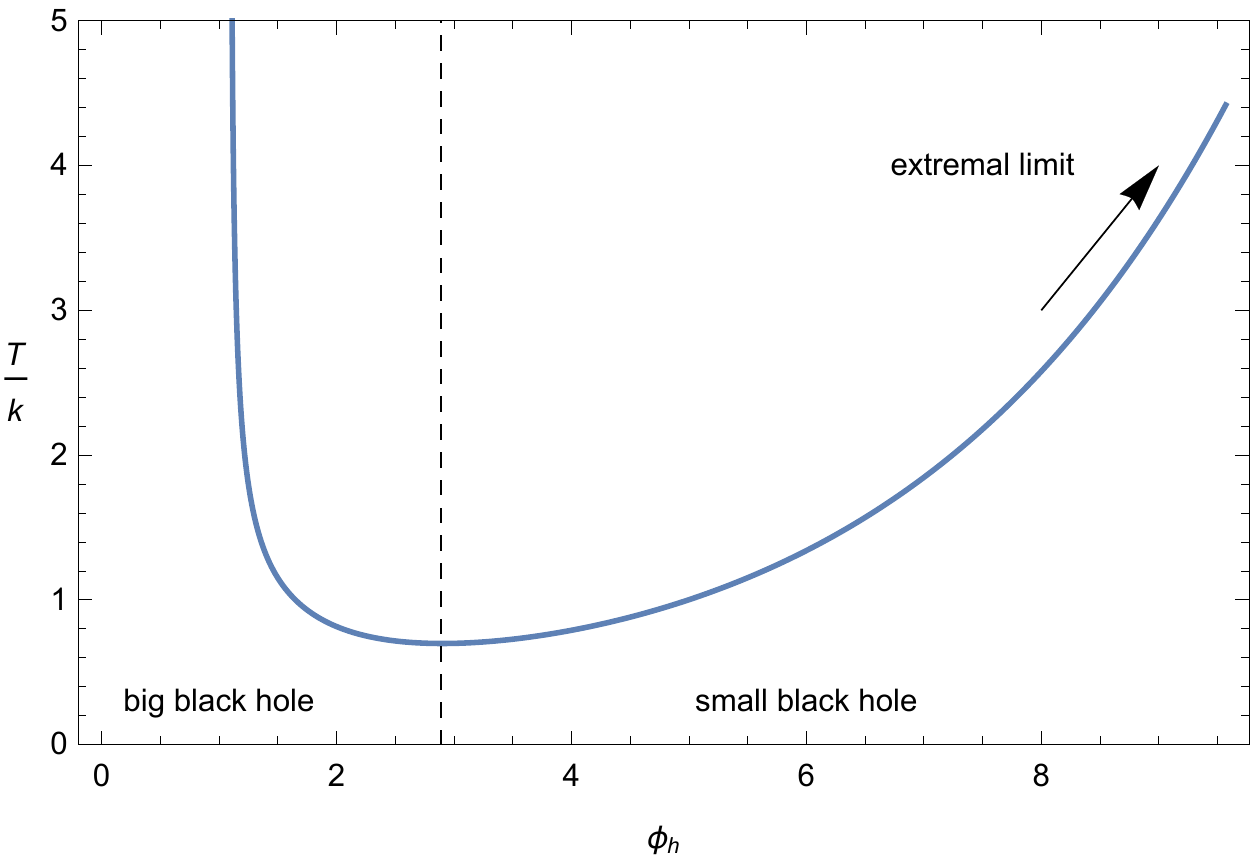}
  \includegraphics[height=140pt]{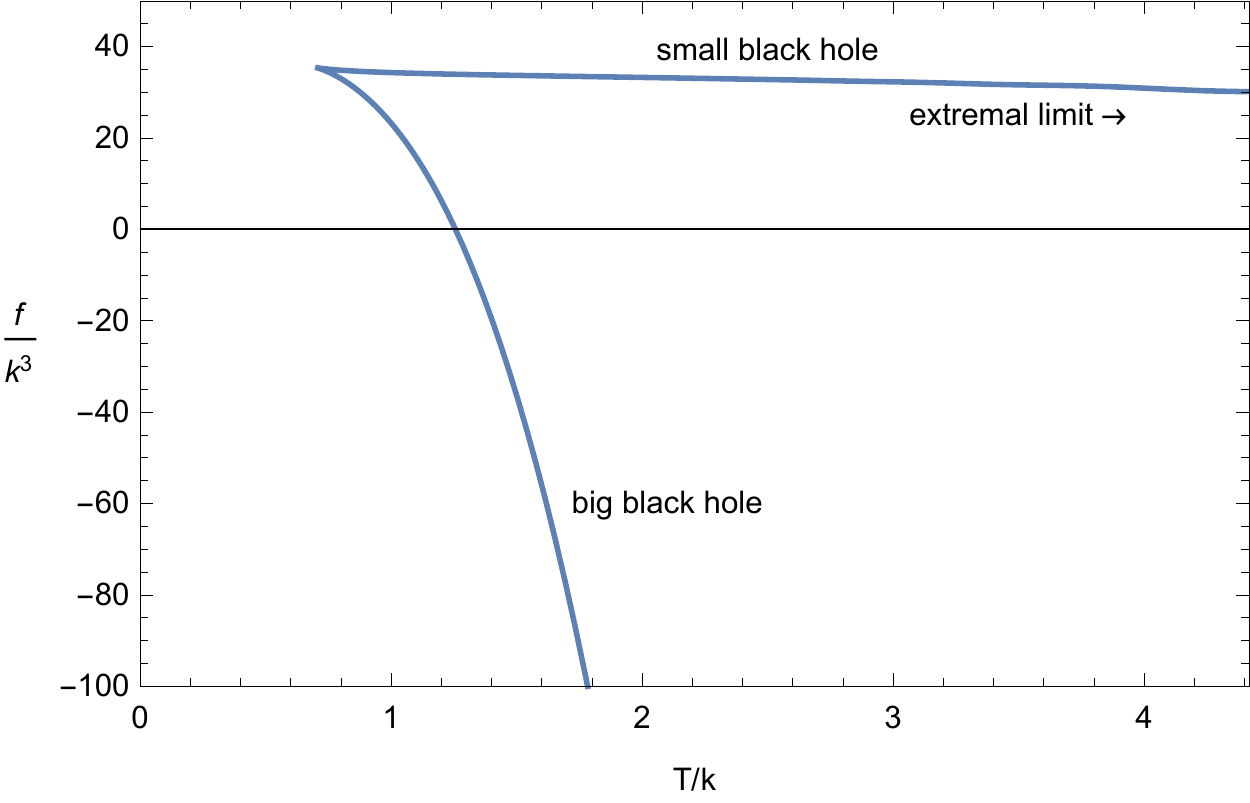}
  \includegraphics[height=140pt]{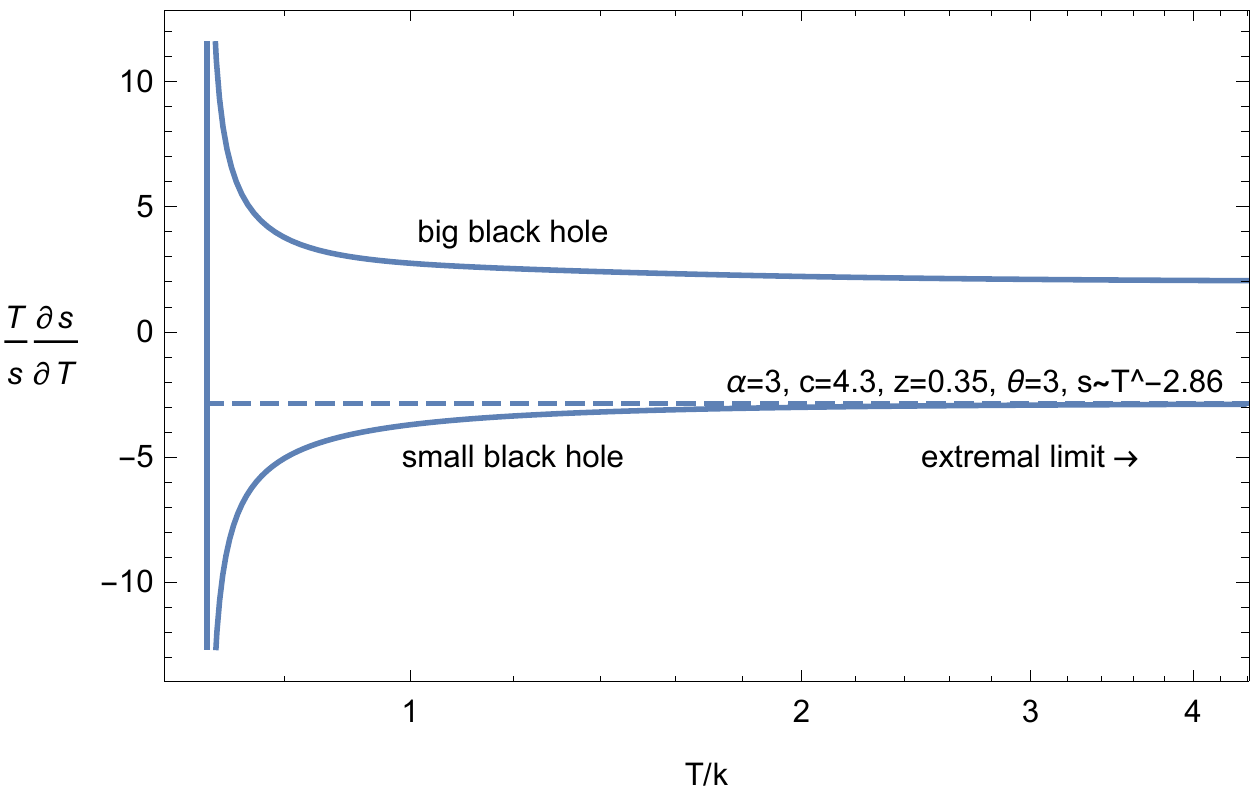}
  \includegraphics[height=140pt]{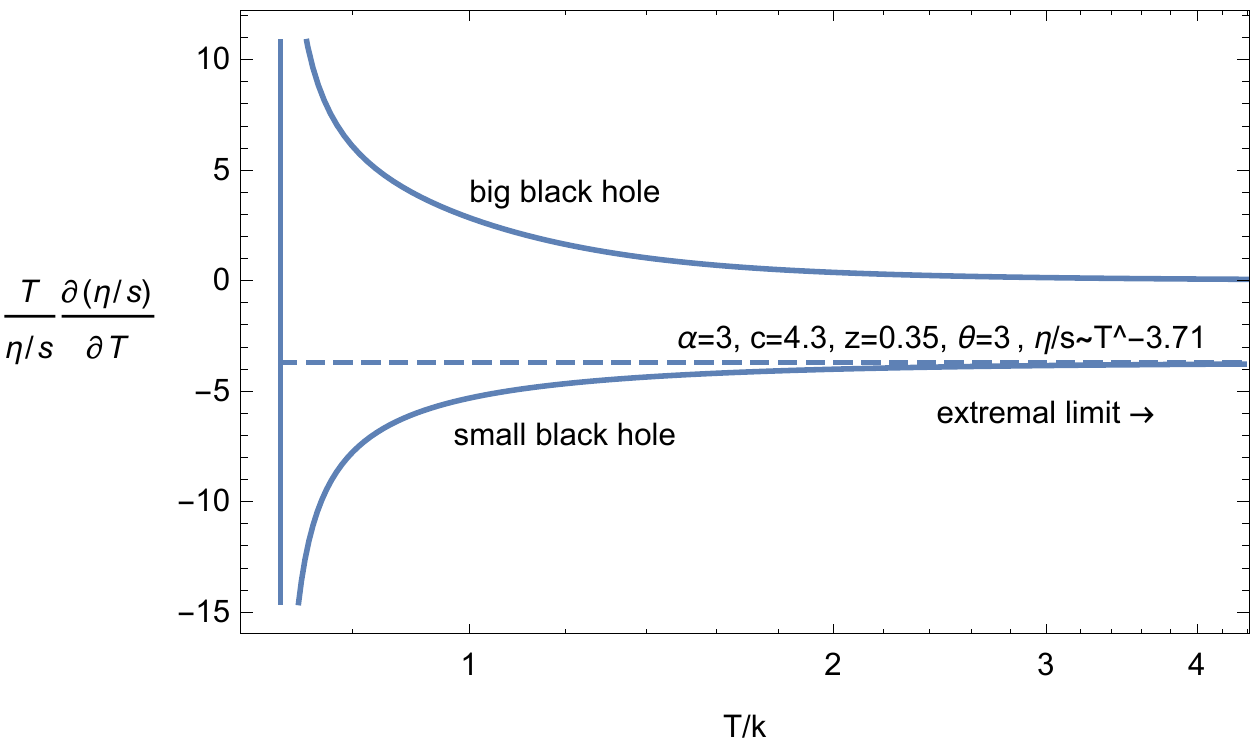}
  \caption{Numerical results in Region B with $\alpha=3$ and $\,c=4.3$. The left-upper plot shows $T/k$ as a function of $\phi_h=\phi(1)$. The right-upper plot shows the dimensionless free energy density $f/k^3$ as a function of $T/k$. The scaling exponents of $s\sim T^\lambda$ (the left-lower plot) and $\frac{\eta}{s}\sim T^\kappa$ (the right-lower plot) as a function of $T/k$. The extremal limit is at $T\to+\infty$ for small black hole.
  }\label{TpowersA2}
\end{figure}

\begin{itemize}
  \item Figure \ref{TpowersA1} is a typical plotting for the temperature behavior of $s$ and $\eta/s$ in Region A. At low temperature, the scaling exponents of $s\sim T^\lambda$ and $\frac{\eta}{s}\sim T^\kappa$ through numerical calculation match the analytical results from Eq. (\ref{EMDAscaling}) very well. In particular, the values of exponent $\kappa$ are greater than 2, in contrast to the results in \cite{Hartnoll:2016tri}. At high temperature, the numerical results approach to $s\sim T^2,\,\frac{\eta}{s}\sim 1$, which is the standard result for the usual AdS-Schwarzschild black hole.

  \item Figure \ref{TpowersA2} is a typical plotting for the temperature behavior of $s$ and $\eta/s$ in Region B. In the left-upper plot of Figure \ref{TpowersA2}, we notice that above the minimal temperature $T_{min}$, there are two branches of black hole solutions, one corresponding to  big black holes while the other to  small black holes \cite{Gursoy:2008za,Gursoy:2008bu,Kiritsis:2015oxa}. The $T\to+\infty$ limit of the big black hole is the AdS-Schwarzschild black hole. The extremal limit can be approached by heating the small black hole to $T\to+\infty$.

      The small black hole branch is thermodynamically unstable as expected, since its free energy density is greater than the one in big black hole branch with the same temperature, as shown in the right-upper plot of figure \ref{TpowersA2}. Above certain a critical temperature $T_c>T_{min}$, the big black hole is thermodynamically dominated. While for $0<T<T_c$, the extremal limit with a periodical time of $t\sim t+iT^{-1}$ is dominated, which is the ground state of the system. A first order phase transition happens at $T_c$ between the ground state and the big black hole branch.

      Anyway, in the extremal limit we find that the scaling exponents from numerical calculation match the one of the hyperscaling violating solution (\ref{EMDAscaling}), as shown in the bottom of Figure \ref{TpowersA2}.
\end{itemize}

In the end of this section we turn to the temperature behavior of $s$ and $\eta/s$ for parameters in region C, in which the choice for UV completion is different. Since $\alpha<0$, if we expect that the term of $e^{\alpha\phi}$ in potential $V(\phi)$ is leading in the IR region, we need $\phi\to-\infty$. Thus we have to choose $J(\phi)=\frac{c}{4}\,\text{sech}^2(\phi)$ in order to reach the right hyperscaling violating solution above. Consequently, the modified  action with UV completion for Region C is
\begin{equation}\label{UVactionA3}
\mathcal{S}=\int dt d^2x dr\sqrt{-g}\left\{ R+ 6 + \frac{4c}{\alpha^2}\sinh(\frac{\alpha}{2}\phi)^2 - \frac{c}{2}\left[(\partial\phi)^2 + \frac14\,\text{sech}^2(\phi) \sum_{i=1,2}(\partial\chi_i)^2 \right]\right\}.
\end{equation}

It should be noticed that though as $\phi\to-\infty$, $\text{sech}^2(\phi)\to0$, we still have $\text{sech}^2(\phi)(\partial\chi_i)^2\sim r^{-\alpha}\to+\infty$ to build up the ground state with relevant lattices.

Besides $AdS_4$ vacuum with unit radius, the action also admits a solution with extremal geometry of $AdS_2\times R^2$ and vanishing dilaton
\begin{equation}
ds^2=\frac13ds^2_{AdS_2}+dx_1^2+dx_2^2,\quad \chi_i=\sqrt{\frac{24}{c}}x_i, \quad \phi=0.
\end{equation}
Since $\frac{c}{8}\,\text{sech}^2(\phi)\sum_i(\partial\chi_i)^2=6-6\phi^2+\cdots$, the square of effective mass of dilaton is $-(2+12/c)$, which violates the $AdS_2$ BF bound of $-\frac34$. We expect that a condensation of dilaton occurs at relatively low temperature (but it is still at high temperature with respect to the emergence of hyperscaling violation).

Respect to the $AdS_4$ vacuum, the square of mass of dilaton is $-2$ as well. Here we choose the conformal weight of the dual operator of dilaton as $\Delta=2$ and demand its source to be zero by choosing one of the AdS boundary conditions as $\phi'(0)=0$ for numerical convenience. The other boundary conditions are the same as those in Region A and B. Thus there is only one parameter $T/k$ remaining.

\begin{figure}
  \centering
  \includegraphics[width=230pt]{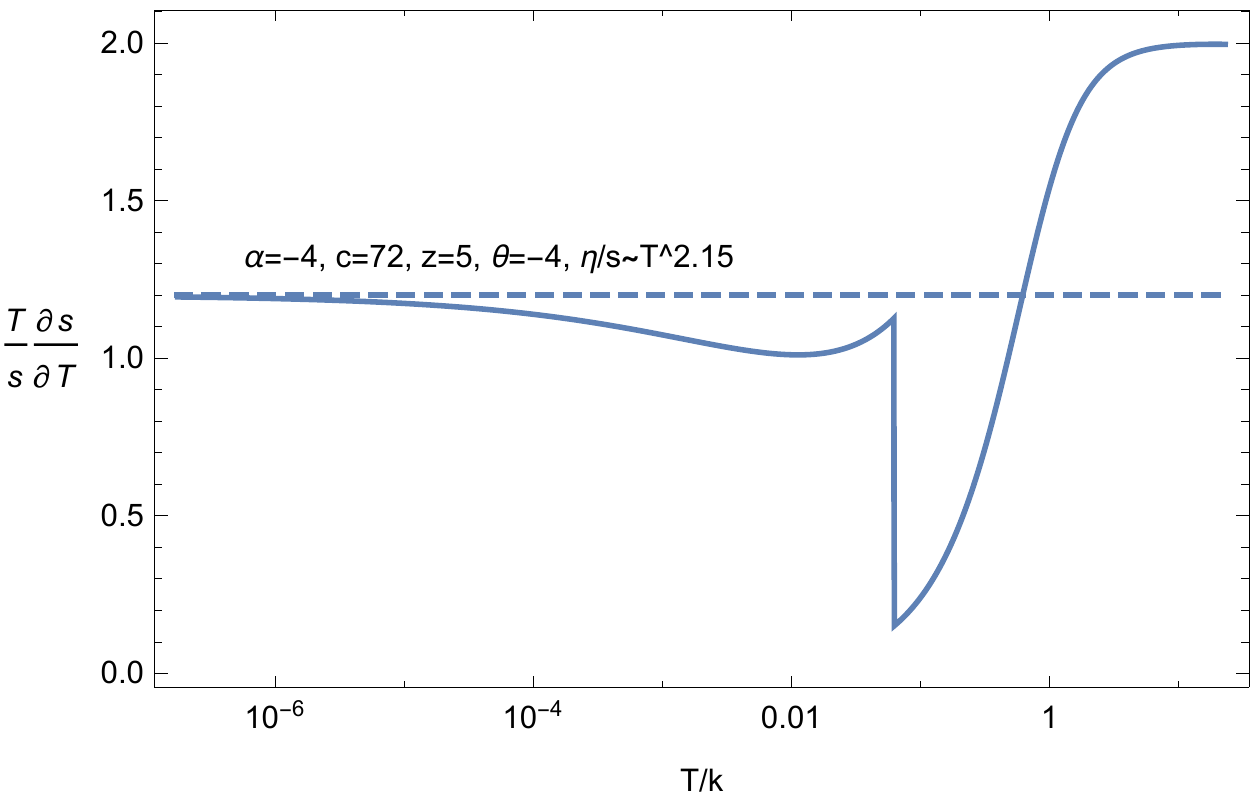}
  \includegraphics[width=230pt]{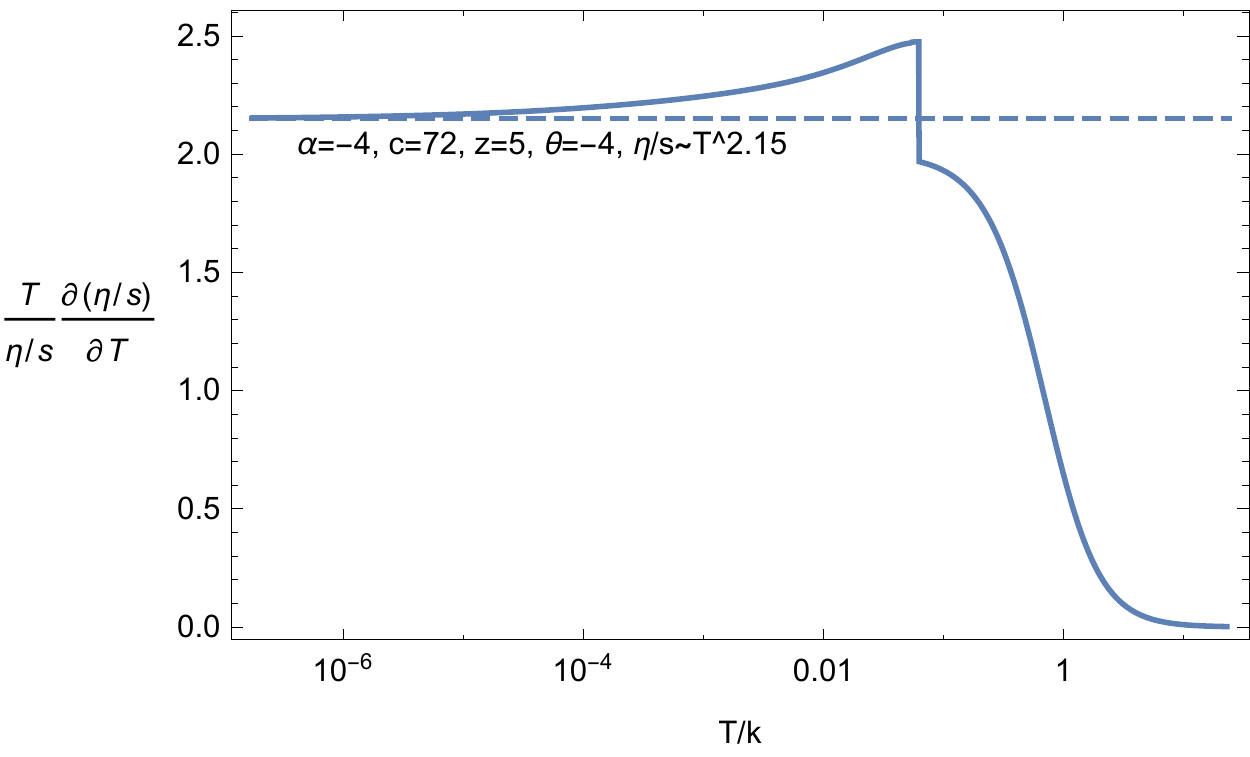}
  \caption{The scaling exponents of $s\sim T^\lambda$ (the left plot) and $\frac{\eta}{s}\sim T^\kappa$ (the right plot) as a function of $T/k$ in Region C. The step at $T/k\approx0.06$ results from the second order phase transition. The extremal limit is at $T\to0$.}\label{TpowersA3}
\end{figure}

We numerically find that the dilaton condensates spontaneously at
relatively low temperature. It leads to a second order phase
transition between pure axion black hole and dilaton-axion black
hole. By comparing the free energy, we find that the dilaton-axion
black hole is thermodynamically dominated below the
critical temperature. When we continuously drop down the
temperature, the hyperscaling violating solution is approached and
the power law is verified, as shown in Figure \ref{TpowersA3}. We
can see that $\kappa$ tends to the predicted number which is
greater than 2.

\section{Isotropic dilaton-linear axion lattices with infinite $(z,\theta)$: $\eta$-geometry}\label{Sectionetageometry}
By rescaling the dilaton and parameters of (\ref{IRactionA}) as in \cite{Donos:2014uba}, we obtain the following action.
\begin{equation}\label{IRactionB}
\mathcal{S}=\int dt d^2x dr\sqrt{-g}\left[ R - \frac12(\partial\phi)^2 - \frac12\sum_{i=1,2}(\partial\chi_i)^2 +n_1e^{\alpha\phi}\right],
\end{equation}
where we have chosen the potentials as $J(\phi)=1 ,\, V(\phi)=n_1e^{\alpha\phi}$. It has a ground state which is conformal to $AdS_2\times R^2$ with lattices.
\begin{equation}\begin{split}\label{etageometry}
&
ds^2=\frac1{r^{\bar{\eta}}}\left( L^2\frac{-dt^2+dr^2}{r^2}+dx_1^2+dx_2^2 \right), \, e^\phi=r^{\phi_1}, \, \chi_{1,2}=kx_{1,2},
\\&
\bar{\eta} =\frac{2 \alpha ^2}{1-\alpha ^2},\,k^2=n_1(1-\alpha ^2),\,L^2=\frac{2 (1+\alpha ^2)}{n_1(1-\alpha ^2)^2},\,\phi _1=\frac{2 \alpha }{1-\alpha ^2}.
\end{split}\end{equation}
It corresponds to the situation of $z,\theta\to\infty$ while keeping $-\frac{\theta}{z}=\bar{\eta}$ fixed.

From $L^2>0, k^2>0$, we have $n_1>0,\alpha^2<1,\bar{\eta}>0$ and
$r\xrightarrow{IR}+\infty$. We just choose $0<\alpha<1$ such that
$\phi_1>0$ and $\phi\to+\infty$ at IR. Since $\bar\eta>0$, here
the $\eta$-geometry can be obtained by taking the limit of
$(z\to\mp\infty,\theta\to\pm\infty)$ in Region A or Region C.

Applying the following mode analysis
\begin{equation}
ds^2=\frac1{r^{\bar{\eta}}}\left[ L^2\frac{-(1+c_tr^\delta)dt^2+(1+c_rr^\delta)dr^2}{r^2}+(1+c_xr^\delta)(dx_1^2+dx_2^2) \right], \, e^\phi=r^{\phi_1}(1+c_\phi r^\delta),
\end{equation}
we find modes which are similar to Section \ref{Sectionlatticesgeometry}. There are two pairs of modes which satisfy $\delta_-+\delta_+=1+\bar{\eta}$ after getting rid of the trivial modes \footnote{The trivial modes are proportion to $c_t=-\eta-2,c_r=-\eta+2\delta,c_x=-\eta,c_\phi=\phi_1$ which correspond to infinitesimal transformation $r\to r(1+\epsilon r^\delta)$ for any $\delta$, where $\epsilon<<1$.}. One pair has $\delta^{(0)}_-=0$ and $\delta^{(0)}_+=1+\bar{\eta}$ ($-c_t=c_r=r_+^{-1-\eta},c_x=c_\phi=0$), which correspond to rescaling of time and creating a small black hole with temperature $T\propto r_+^{-1}$. The other pair has
\begin{equation}
\delta^{(1)}_\pm=\frac{1}{2} \left(1+\bar{\eta} \pm\sqrt{(1+\bar{\eta}) (9+\bar{\eta})}\right),
\end{equation}
which satisfies $\delta^{(1)}_-\delta^{(1)}_+=-2(1+\bar{\eta})<0$. Then the scaling solution above are RG stable, for the same reason in Section \ref{Sectionlatticesgeometry}.

As pointed out in Section \ref{SectionHVlatticescale}, the power law of $s$ and $\eta/s$ reads separately as
\begin{equation}
s\sim T^{\bar{\eta}},\quad \frac{\eta}{s}\sim T^{(1+\bar{\eta}) \left(-1+\sqrt{1+\frac{8}{1+\bar{\eta}}}\right)}.
\end{equation}
From $\bar{\eta}>0$, we have $2<\kappa<4$.

We adopt the following UV completing action.
\begin{equation}\label{UVactionB}
\mathcal{S}=\int dt d^2x dr\sqrt{-g}\left\{ R+ 6 + \frac{4}{\alpha^2}\sinh(\frac{\alpha}{2}\phi)^2 - \frac12 (\partial\phi)^2 - \frac12 \sum_{i=1,2}(\partial\chi_i)^2  \right\}.
\end{equation}
Similar to Region C in Section \ref{Sectionlatticesgeometry}, the action admits $AdS_4$ vacuum with unit radius and $AdS_2\times R^2$ with radius of $1/\sqrt{3}$. The square of mass of dilaton is $-2$ and violates the $AdS_2$ BF of $-\frac34$. We expect a condensation of dilaton.

\begin{figure}
  \centering
  \includegraphics[width=230pt]{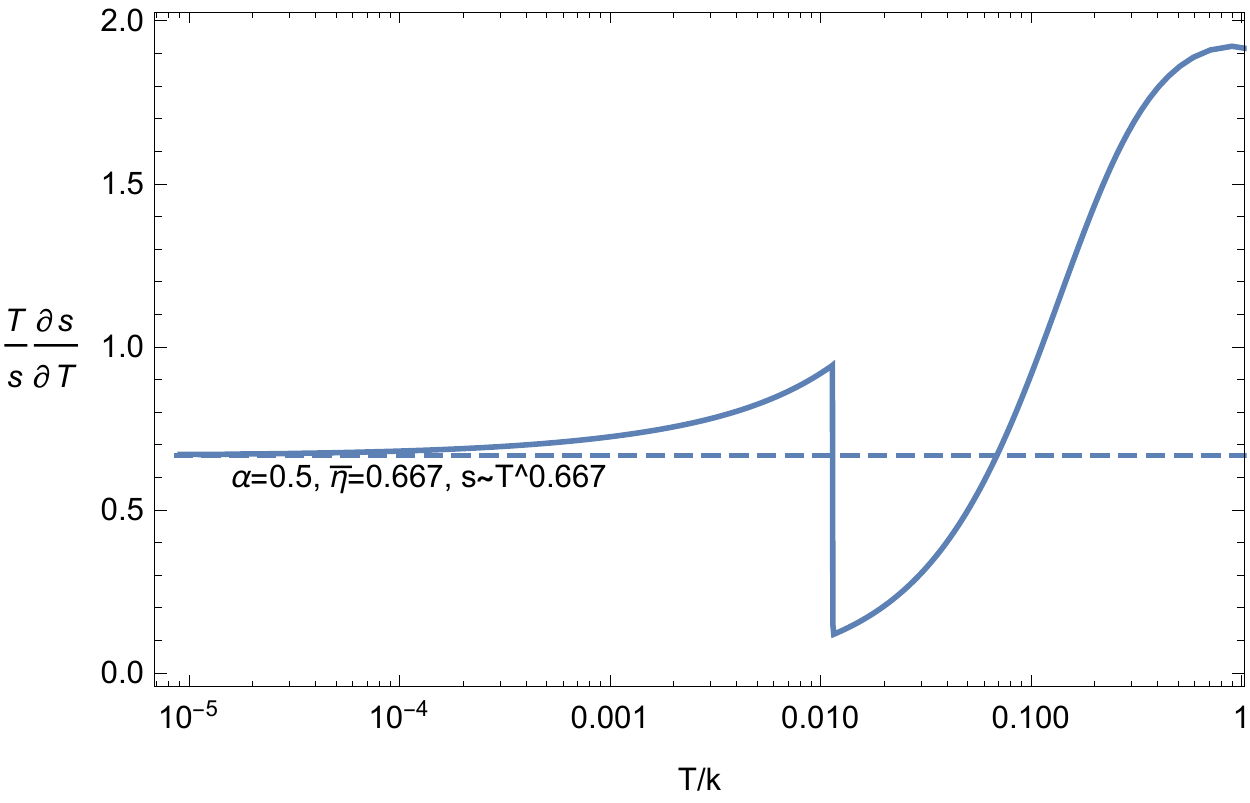}
  \includegraphics[width=230pt]{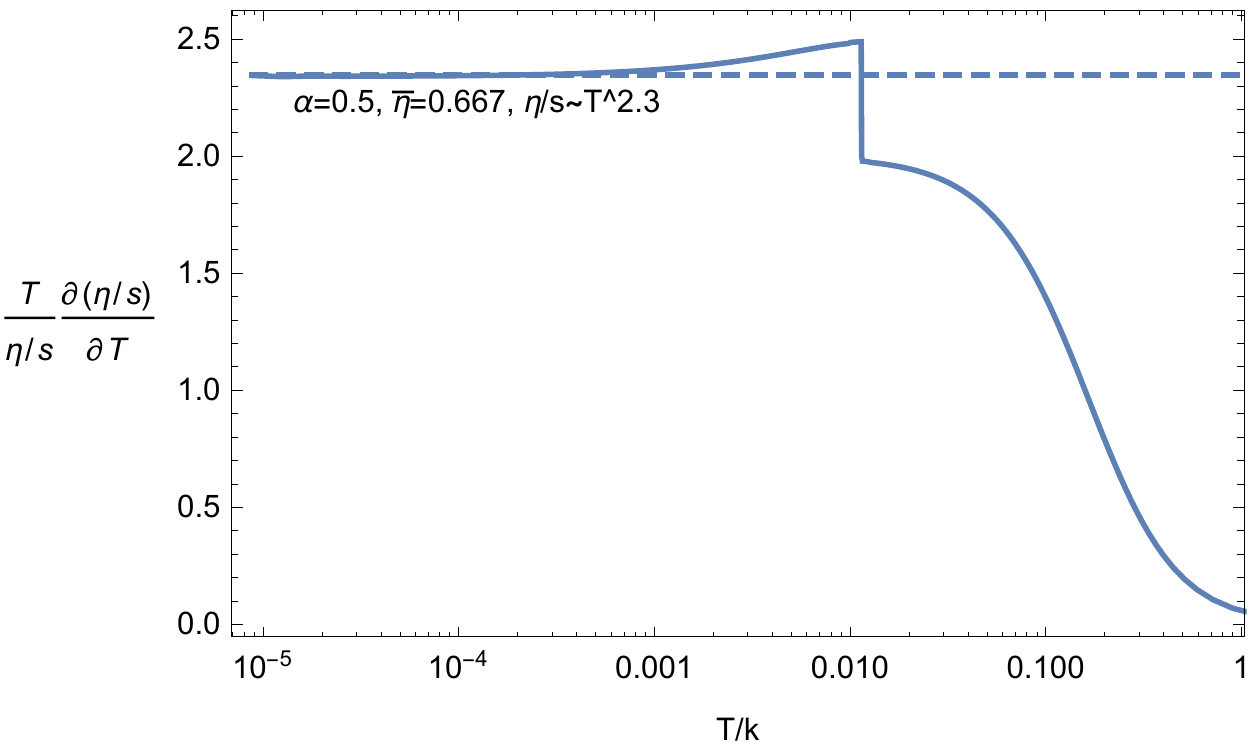}
  \caption{The scaling exponents of $s\sim T^\lambda$ (the left plot) and $\frac{\eta}{s}\sim T^\kappa$ (the right plot) as a function of $T/k$. The step at $T/k\approx0.01$ results from the second order phase transition between pure axion black hole and dilaton-axion black hole. The extremal limit is at $T\to0$.}\label{TpowersB}
\end{figure}

Ansatz and the boundary conditions for numerical calculation are chosen to be the same as those in Region C. The hyperscaling violating solution is approached at low temperature and the power law is verified, as shown in Figure \ref{TpowersB}.

\section{Discussion and outlooks}\label{Sectiondiscussion}
\subsection{In comparison with the behavior of entanglement entropy}
We have constructed specific models with the violation of
the shear viscosity bound (\ref{etaspowerlaw}) in Section
\ref{Sectionlatticesgeometry} and \ref{Sectionetageometry}, which
has been verified by numerical calculation. It becomes
urgent to understand the underlying reasons leading to such
violations. Apparently, the violation might be rooted in the
nonzero exponent of hyperscaling violation $\theta$. But our
analysis for higher dimension $d>2$ indicates that such kind of
violation can occur even $\theta=0$, as shown in Figure
\ref{boundvszd}.

As mentioned in the end of Subsection \ref{SubSectionHVmetirc}, investigations of the behaviors of entanglement entropy give further constraint to the hyperscaling violating theories. We suggest that the bound violation may be related to the peculiar behavior
of entanglement entropy in these theories. Explicitly,
\begin{itemize}
  \item When $d_\text{eff}>1$ (containing Region C), we find $0\leq\kappa<4$,  suggesting a new bound of 4 rather than 2. Within this region, the entanglement entropy is subject to the area law, implying that the dual local QFTs do not have large accidental degeneracies in low energy spectrum \cite{Huijse:2011ef}. In addition, for $\eta$-geometry (\ref{etageometry}), the upper bound is 4 and the entanglement entropy satisfies the area law as well, under the condition that the width of the strip is large enough \cite{Liu:2013una}.
  \item When $0\leq d_\text{eff}\leq 1$, we find $\kappa\leq2$ from the power law (\ref{bound}), which just coincides with the bound (\ref{etaspowerlaw}). Within this region, the area law of entanglement entropy receives violations interpolating between the logarithmic and linear behavior \cite{Dong:2012se,Huijse:2011ef}. Especially, when $d_\text{eff}=1$, a logarithmic violation appears, signaling the existence of fermi surface; when $d_\text{eff}=0$, a linear violation appears and leads to a volume law, signaling an extensive ground state entropy. Recall that the known extremal IR geometries with nonvanishing lattices studied in \cite{Hartnoll:2016tri} are $AdS_2\times R^2$, which belong to the case of $d_\text{eff}=0$, and the entanglement entropy shows volume law.
  \item When $d_\text{eff}<0$, we find $\kappa>2$ for $z<0 \land e=0$ (containing Region A) and $\kappa\leq0$ for $0<z\leq1$ (containing Region B), which just violate the bound (\ref{etaspowerlaw}), neither more nor less, and suggesting the inexistence of the bound. Within this region, the entanglement entropy scales faster than the volume, which is not the behavior of QFT. Moreover, the stationary surface of entanglement entropy becomes a maximum, which suggests some instability of gravitational background \cite{Dong:2012se}. So the violation of $0<\kappa<2$ in this region might be related to the abnormality of entanglement entropy and gravitational background.
\end{itemize}

From the analysis above, we give a conjecture that the bound of
$\kappa$ depends on the behaviors of entanglement entropy, due to
the different natures of ground states: when entanglement entropy
shows area law, the bound is 4; when the area law have logarithmic
to linear violation, the bound is 2; when the volume law is
exceeded, then there is no bound.

\subsection{Conclusions and open questions}
In this paper we have investigated the shear viscosity in a
general holographic framework with hyperscaling violation. In the
presence of isotropic and relevant lattices, we have demonstrated that the
scaling relation in extremal IR region strongly constrains the
mass term of graviton such that the ratio of shear viscosity to
the entropy density always exhibits a power law behavior with
temperature, $\eta/s\sim T^\kappa$. Significantly, we have found
that in the EMD-Axion theory (\ref{action}) the exponent $\kappa$
can be greater than 2 such that the bound (\ref{etaspowerlaw})
raised in \cite{Hartnoll:2016tri} is violated. Our above
observation has been verified by numerically constructing a large
class of black hole solutions with UV completion in the EMD-Axion
theory. On the other hand, when the axion is irrelevant, at subleading order, $k$
appears in the expression of $\eta/s$ as another scale and leads
to a complicated behavior of temperature dependence (\ref{h0hirrelevant}) which is
beyond the simple power law.

It is instructive to discuss the bound of entropy production
rate in the holographic framework with hyperscaling violation,
closely following the consideration presented in
\cite{Hartnoll:2016tri}. As analyzed in Section
\ref{Sectionscalingofeta}, when breaking of translational invariance is relevant in the IR, operate $\hat{T}^{xy}$ acquires
a scaling dimension of $\delta_{\hat T}$ in the IR, so as its dual
source $\delta g_{xy}^{(0)}$ acquires
$\delta_0-\delta_{\hat T}$, denoted as $[\delta
g_{xy}^{(0)}]=\delta_0-\delta_{\hat T}$. Consider the source
$\delta g_{xy}^{(0)}$ to be linear in time as proposed in
\cite{Hartnoll:2016tri}
\begin{equation}
\delta g_{xy}^{(0)}= t c^{(0)},
\end{equation}
where $c^{(0)}$ is a time independent constant. Since $[t]=-[T]=-z$,
then $[c^{(0)}]=\delta_0-\delta_{\hat T}+z$. On the other hand, from Eq.(\ref{viscosityscaling}), we have
$[\eta/s]=2(\delta_{\hat T}-\delta_0)$. Then the equation about
the rate of entropy density production
\begin{equation}
\frac1{T}\frac{d\log
s}{dt}=\frac{\eta}{s}\left(\frac{c^{(0)}}{T}\right)^2,
\end{equation}
has scaling dimensions of zero on both sides, which
is natural. Then the bound of entropy production
rate is still allowable,
\begin{equation}
t_\text{Pl}\frac{d\log(s)}{dt}\gtrsim 1,
\end{equation}
where $t_\text{Pl}=\frac{\hbar}{k_BT}$ is the `Planckian time'. Let us assume that
temperature $T$ is still a dominating scale. Then $c^{(0)}=T^{\frac{\delta_0-\delta_{\hat T}}{z}+1}$
is the natural choice which satisfies the scaling
dimension\footnote{Our strategy here is
different from that in \cite{Hartnoll:2016tri}, where
it is argued that the strain constant $c^{(0)}$ can be another scale surviving in the
IR besides temperature $T$, such as momentum scale. We thank Sean
Hartnoll for helpful suggestions.}.

We have conjectured that the boundedness of $\kappa$ relates to
the behavior of entanglement entropy. In particular, when the area
law of entanglement entropy is satisfied, a higher bound of 4 for
$\kappa$ has been suggested. The reason of the boundedness
of $\kappa$  might be ascribed to the boundedness of scaling
dimension of operator $\hat{T}^{xy}$.

Finally, a lot of open problems deserve for further investigation.
Firstly, in this paper we have only considered the isotropic
lattices due to the axion fields. In Section
\ref{Sectionlatticesgeometry} and \ref{Sectionetageometry}, we
only do the calculation at the scaling solutions with vanishing
current, and the UV complete solutions with (marginally) relevant
current and (marginally) relevant lattices are worthy of
investigation in future \cite{Gouteraux:2014hca,Ling:2016yxy}.
On the other hand, the anisotropic situation is interesting as well, since an
anisotropic scaling relation will emerge in the IR
\cite{Donos:2014uba,Donos:2014oha}. Furthermore, by defining an
effective (scaleless) mass of graviton, we expect that our scaling
analysis on shear viscosity can be generalized to models in which
the translational symmetry is broken by other effects, such as
massive gravity
\cite{Vegh:2013sk,Andrade:2013gsa,Alberte:2016xja}, magnetic
charge \cite{Liu:2016njg} or disordering
\cite{Hartnoll:2014cua,Hartnoll:2015faa,Hartnoll:2015rza}, since
the scaling relations emerged in the IR belong to one sort of
hyperscaling violations.

Secondly, since other components of graviton are massive as well,
Green functions associated with other components of
energy-momentum tensor may exhibit similar scaling behaviors, then
their susceptibilities, such as bulk viscosity, are expected to
exhibit some power laws of temperature.

Thirdly, we stress that it is very crucial to understand the
underlying reasons of boundedness or boundlessness of
$\kappa$ in different regions. One may investigate it from
the viewpoint of dimensional reduction, since the power law to the
temperature may return to a more simple way in higher
dimension. In EMD theories, the solutions of higher-dimensional
theories reducing to $d_\text{eff}<0$ region are asymptotically flat
p-branes \cite{Gouteraux:2011qh,Gouteraux:2011ce}. The
boundlessness of $\kappa$ in this region may come from the absence of the scaling symmetry of AdS or Lifshitz in
higher-dimensional spacetimes, although the exact dimensional
reduction of the EMD-Axion theories is not clear
yet\footnote{We thank Blaise Gout\'eraux for helpful suggestions.}.

The violation of the area law of entanglement entropy
is related to the bound for entanglement entropy production
rate \cite{Acoleyen:2013era}, which has been studied during
thermalization in holographic system \cite{Liu:2013qca,Alishahiha:2014cwa}. The
relation between the shear viscosity bound and entanglement
entropy calls for further investigation.

\begin{acknowledgments}

We are very grateful to Matteo Baggioli, Blaise Gout\'eraux, Sean
Hartnoll, Peng Liu, Mohammad Reza Mohammadi Mozaffar, Diego
Trancanelli, Walter Tangarife and Xiangrong Zheng for helpful
discussions and correspondence. We also thank Gout\'eraux and
Hartnoll for constructive comments on the previous version of our
paper. Finally, we thank the anonymous referee for raising
valuable questions and urging us to complete the content in
Appendix \ref{Appendixirr}. This work is supported by the Natural
Science Foundation of China under Grant Nos.11275208 and 11575195,
and by the grant (No. 14DZ2260700) from the Opening Project of
Shanghai Key Laboratory of High Temperature Superconductors. Y.L.
also acknowledges the support from Jiangxi young scientists
(JingGang Star) program and 555 talent project of Jiangxi
Province.

\end{acknowledgments}

\begin{appendix}

\section{Shear viscosity with (marginally) relevant axion}\label{Appendix}
In this appendix we derive the shear viscosity $\eta$ through
the retarded Green function explicitly. We will show that
the result is consistent with that from the scale analysis
(\ref{viscosityscaling}). We start from the shear
perturbation equation in hyperscaling violating metric
(\ref{ThermalHVmetric}) which reads as
\begin{equation}
\partial_r(r^{1-\delta_0}f(r)\partial_r h(r))+(\frac{r^{2z-\delta_0-1}\omega^2}{f(r)}-M^2L^2 r^{-\delta_0-1} ) h(r) =0,\label{peq}
\end{equation}
where
\begin{equation}
f(r)=1-\left(\frac{r}{r_+}\right)^{\delta_0},~~~~\delta_0:=d+z-\theta.
\end{equation}
Note that we have used $m(r)^2=M^2 r^{-\frac{2 \theta }{d}}$ as discussed in Section \ref{Sectionscalingofeta}. We remind that $M^2\geq0,L^2>0$ and temperature is $T=|\delta_0| r_+^{-z}/(4\pi)$.

To solve this equation, we change it into a transparent form by defining
\begin{equation}
\xi:=\frac{r^{\delta_0}}{r_+^{\delta_0}}\,,~~~~~a:=\frac{1}{2}\left(1-\sqrt{1+\left(\frac{2ML}{\delta_0}\right)^2}\right)\label{newv},
\end{equation}
where $a\leq0$. The new coordinate $\xi$ covers the region $1\geq\xi\geq0$, with the horizon at $\xi=1$ and the boundary at $\xi=0$.
Now, the perturbation equation (\ref{peq}) can be rewritten as
\begin{eqnarray}
(1-\xi)\partial^2_\xi h(\xi)-\partial_\xi h(\xi)+\left(\frac{\xi^{\frac{2z}{\delta_0}-2}}{1-\xi}\left(\frac{\omega}{4\pi T}\right)^2-\frac{a(a-1)}{\xi^2} \right) h(\xi) =0.\label{peq1}
\end{eqnarray}
As we will see below, the term of $\omega^2$ is not important for
calculating the viscosity. With regularity condition at horizon,
the zero frequency solution $h_0(\xi)$ can be obtained as
\begin{eqnarray}
h_0(\xi)= \, \xi^a \, _2F_1(a,a;2 a;\xi)-\frac{\Gamma (1-a)^2
\Gamma (2 a)}{\Gamma (2-2 a) \Gamma (a)^2}\, \xi^{1-a} \,
_2F_1(1-a,1-a;2-2 a;\xi),\label{0sol}
\end{eqnarray}
where $_2F_1(a,b;c;z)$ is the Gaussian hypergeometric function.
Especially, at the horizon we have
\begin{eqnarray}
h_0(1)= \frac{\pi ^2 \csc ^2(\pi  a)}{\Gamma (1-2 a) \Gamma (a)^2},\label{hh}
\end{eqnarray}
while on the boundary, $h_0(\xi)$ behaves as
\begin{eqnarray}\label{hb}
h_0(\xi\rightarrow 0)=\xi^a+\cdots -\frac{\Gamma (1-a)^2 \Gamma (2 a)}{\Gamma (2-2 a) \Gamma (a)^2}\, \xi^{1-a}+\cdots,
\end{eqnarray}
which is the explicit form of (\ref{boundarybehavior}).

We next introduce the in-falling boundary condition and expand the solution in power of the frequency as
\begin{equation}\begin{split}
h(\xi)&=(1-\xi)^{\frac{- i\omega}{4\pi T}}h_0(\xi)(1+i \omega H(\xi)+\mathcal{O}(\omega)^2)\\
&=h_0(\xi)(1+i\omega \tilde{H}(\xi)+\mathcal{O}(\omega)^2),
\end{split}\end{equation}
where $H(\xi)$ is regular at the horizon and $\tilde{H}(\xi)=H(\xi)-\frac{\ln(1-\xi)}{4\pi T}$.
Then, substituting the above expansion into (\ref{peq1}), we derive a conservation equation up to the first order of $\omega$
\begin{eqnarray}
\partial_\xi(h_0(\xi)^2(1-\xi)\partial_\xi\tilde{H}(\xi)) =0.
\end{eqnarray}
Now we evaluate the conserved quantity $h_0(\xi)^2(1-\xi)\partial_\xi\tilde{H}(\xi)$ at the horizon, leading to
\begin{eqnarray}
\partial_\xi\tilde{H}(\xi)=\frac{h_0(1)^2}{4\pi T (1-\xi)h_0(\xi)^2}.
\end{eqnarray}
This result gives the asymptotic behavior of $\tilde{H}(\xi)$ on the boundary
\begin{eqnarray}
\tilde{H}(\xi\rightarrow 0)=C+\frac{h_0(1)^2}{4\pi T (1-2a)}\xi^{-2a+1}+\cdots,
\end{eqnarray}
where (\ref{hb}) has been used and $C$ is an integration  constant.
Finally, we have the  asymptotic behavior of $h(\xi)$ on the boundary
\begin{eqnarray}\label{hxiomega}
h(\xi\rightarrow0)=\xi^a+\cdots+\left(i\omega\frac{h_0(1)^2}{(1-2a)4\pi T}-\frac{\Gamma (1-a)^2 \Gamma (2 a)}{\Gamma (2-2 a) \Gamma (a)^2}\right)\xi^{-a+1}+\cdots+\mathcal{O}(\omega^2).
\end{eqnarray}
Next we derive the viscosity from the imaginary part of the retarded Green function in (\ref{boundarybehavioromega}). To do that we change the coordinate $\xi$ in (\ref{hxiomega}) back to the original one in (\ref{newv}). We find the viscosity takes the following form in hyperscaling violating geometry
\begin{eqnarray}\label{viscosityHV}
\eta_\text{HV}=\lim_{\omega\to0}\frac{\text{Im}\, \mathcal{G}^R_{\hat{T}^{xy},\hat{T}^{xy}}(\omega,k=0)}{\omega}
=\frac{h_0(1)^2}{b|\delta_0|(1-2a)}\left(\frac{4\pi T}{|\delta_0|}\right)^{-1+\frac{\delta_0}{z}\sqrt{1+\left(\frac{2ML}{\delta_0}\right)^2} },
\end{eqnarray}
where (\ref{TsHV}) has been used and $h_0(1)$ is given by (\ref{hh}).

On the other hand, given the entropy density in
(\ref{TsHV}), thus we have
\begin{eqnarray}\label{KSSboundHVb}
\frac{\eta_\text{HV}}{s}=\frac{h_0(1)^2}{4\pi b|\delta_0|(1-2a)}\left(\frac{4\pi T}{|\delta_0|}\right)^{\frac{\delta_0}{z}\left(-1+\sqrt{1+\left(\frac{2ML}{\delta_0}\right)^2}\right) }.
\end{eqnarray}
By using the UV-IR matching explained in Section \ref{Sectionscalingofeta}, we have $\frac{\eta}{s}\propto\frac{\eta_\text{HV}}{s}$. Therefore, our result obtained from Green function confirms the temperature behavior of $\eta/s$ given by the scaling analysis (\ref{viscosityscaling}). Besides, if hyperscaling violation is also valid in the UV, i.e. without the AdS-UV completion, the holographic renormalization for a certain hyperscaling violating theory is needed, and the constant $b$ in the expansion (\ref{boundarybehavioromega}) could be determined, at least for EMD theory \cite{Chemissany:2014xpa,Taylor:2015glc}. While, for EMD-Axion theory, the holographic renormalization may be very different, since the scaling dimension of $\hat{T}^{xy}$ is $\delta_{\hat T}$ now, which can deviate from the usual value of $\delta_0=d-\theta+z$ in the translational invariance cases \cite{Chemissany:2014xpa,Taylor:2015glc}.

\section{Shear viscosity with irrelevant axion}\label{Appendixirr}
We are going to study $\eta/s$ on EMD-Axion model with irrelevant
axion at subleading order. We will consider the
perturbation of $h_0(r)$ and find the solution up to ${\cal
O}(k^2)$ and then derive the temperature dependence of
$\eta/s$. Finally, we come to numerical calculation to
justify our formula.
\subsection{Analytical consideration and approximation}
We rewrite the action of EMD-Axion model in which the
hyperscaling violation is allowable \cite{Gouteraux:2014hca}
\begin{equation}
\mathcal{S}=\int dt d^d x dr\sqrt{-g}\left[ R + V_0 e^{\alpha\phi}
- \frac12(\partial\phi)^2 - \frac12
e^{\beta\phi}\sum_{i=1}^d(\partial\chi_i)^2 -
\frac{e^{\gamma\phi}}{4}F^2\right].
\end{equation}
The black hole solution deformed by irrelevant axion up to
${\cal O}(k^2)$ has the following form
\begin{equation}\begin{split}\label{HVirrelevant}
&
ds^2= r^\frac{2\theta}{d} \left[ -\frac{L_t^2 f(r) dt^2}{r^{2z}}(1+k^2 S_t(r))+ \frac{L_r^2 dr^2}{r^2 f(r)} (1+k^2 S_r(r))+ \frac{L_x^2 \sum_{i=1}^d dx_i^2}{r^2}(1+k^2 S_x(r)) \right],
\\&
A=Qr^{\zeta-z} f(r) dt,\quad e^\phi= e^{\phi_0} r^{\phi_1} (1+k^2 S_\phi (r)), \quad \chi_i=kx_i,
\end{split}\end{equation}
where
\begin{equation}
\phi_1^2=\frac{2(d-\theta)((z-1)d-\theta)}{d}, \quad \alpha\phi_1=-\frac{2\theta}{d}, \quad L_r^2 e^{\alpha\phi_0}=\frac{(d+z-\theta-1)(d+z-\theta)}{n_1}.
\end{equation}
The axion back-reacts to the metric and the dilaton with modes
$S_J(r)$ for
$J=\{t,r,x,\phi\}$. $S_J(r)$ satisfy some nonhomogeneous second order differential equations with sources of axion. Since there are freedom for redefinition of $r$ as in usual mode analysis, we can choose the gauge condition of $S_t(r)=0$. Near the boundary, the asymptotic behaviors of other modes
are $S_J(r)=s_{J,0}r^\Delta+s_{J,1}r^\Delta\left(\frac{r}{r_+}\right)^{\delta_0}+\cdots$, where $\Delta=2+\beta\phi_1$. Axion is irrelevant when $\delta_0\Delta<0$. At the horizon, $S_r(r_+)=S_x(r_+)=S_\phi(r_+)=S_r'(r_+)=0$ are required to eliminate other modes. Then the temperature is $T=\frac{|\delta_0|L_t}{4\pi L_r}r_+^{-z}$.
We have maintained the freedom of rescaling
coordinates into $\{L_t,L_r,L_x\}$ for the convenience of
numerical calculation in the next subsection.

Like the case in
Appendix \ref{Appendix}, we can rewrite the shear perturbation
equation for $\omega=0$ in (\ref{perturh}) with coordinate
$\xi=\left(r/r_+\right)^{\delta_0}$ and solve for $h_0(\xi)$ with $k^2$ expansion
\begin{equation}
h_0(\xi)=h_0^{(0)}(\xi)+k^2 h_0^{(1)}(\xi)+\cdots,
\end{equation}
which are subject to the following iterative equations
\begin{eqnarray}
\partial_\xi((1-\xi)\partial_\xi h_0^{(0)}(\xi))&=&0, \label{h00}\\
\partial_\xi((1-\xi)\partial_\xi h_0^{(1)}(\xi))&=&\frac{L_r^2}{\delta _0^2 L_x^2}e^{\beta\phi_0}r_+^{\Delta}\xi ^{b-2}h_0^{(0)}(\xi )-S_1(\xi ) \partial_\xi h_0^{(0)}(\xi )+S_2(\xi ) \partial_\xi^2 h_0^{(0)}(\xi ), \label{h01}
\end{eqnarray}
where $S_1(\xi)$ and $S_2(\xi)$ are some linear combinations of
$S_J$ and their derivatives and $b=\Delta/\delta_0<0$. We expect
${\cal O}(k^2)$ approximation is enough to fit the low temperature
dependence of $\eta/s$ when $k$ is small. The solution to
Eq.(\ref{h00}) which is regular at the horizon is
$h_0^{(0)}(\xi)=C$, with $C$ being a constant. Plug
it into (\ref{h01}), we find that the terms of $S_1(\xi)$ and
$S_2(\xi)$ vanish and the horizon-regular solution is
$h_0^{(1)}(\xi)=C\frac{ L_r^2 }{\delta _0^2 L_x^2}e^{\beta\phi_0}
r_+^{\Delta }\frac{\xi ^b+b (B_{\xi }(b+1,0)+\log (1-\xi ))}{(b-1)
b}$. So the full solution up to ${\cal O}(k^2)$ is
\begin{equation}\label{h0irrelevant}
h_0(\xi)= C \left[1+K^2 r_+^{\Delta }\frac{ \xi ^b+b (B(\xi;b+1,0)+\log (1-\xi ))}{(b-1)b}\right] \quad \text{with} \quad K^2=\frac{k^2 L_r^2 }{\delta _0^2 L_x^2}e^{\beta\phi_0},
\end{equation}
where $B(\xi;b+1,0)$ is the incomplete beta function whose series
definition is
$B(\xi;b+1,0)=\sum_{i=0}^\infty\frac{\xi^{b+1+i}}{b+1+i}$. Formula
(\ref{h0irrelevant}) can tell us how temperature affects the
value of $h_0(\xi)$ at horizon, namely $h_0(1)$. Let us
focus on the cases in which the extremal limit is at
$T\to0$. At low temperature, hyperscaling violation emerges in the
IR and (\ref{HVirrelevant}) is valid only within an interval
between $r_i$ and $r_+$. It connects to AdS deformed by
matter fields near an intermediate scale $r_i$. The constant $C$
can be determined by evaluating $h_0$ at $r_i$ as
\begin{equation}\label{connect}
h_0(\xi_i)=\Gamma
\end{equation}
where $\xi_i=(r_i/r_+)^{\delta_0}$ and $\Gamma$ can be understood
as the tunnelling rate. Such an idea was proposed
in \cite{Hartnoll:2016tri}, while here we just apply it to the
intermediate scale $r_i$. When temperature is much lower
than other scales, it becomes not important to the RG flow
from AdS to hyperscaling violation. The tunnelling rate,
which characterizes how $h_0$ decays from the conformal
boundary to $r_i$, becomes insensitive to temperature and
is expected to go to a constant at low temperature. So temperature
mainly controls $h_0$ by varying $r_+$ in (\ref{h0irrelevant}).
Although we have no general analytical solution with UV completion
and can not determine $C,\,\Gamma$ or $r_i$ analytically, we can
estimate them by numerical fitting in the next subsection.

By working out $C$ from (\ref{connect}), we obtain the value of
$h_0$ at horizon as
\begin{eqnarray}
h_0(\xi=1)&=&\Gamma\frac{  (b-1)b-K^2 r_+^{\Delta }\left(b H_b-1\right)}{(b-1)b + K^2 r_+^{\Delta } \left[\xi _i^b+b \left(B_{\xi _i}(b+1,0)+\log \left(1-\xi _i\right)\right)\right]} \label{h0hirrelevant}\\
&\approx& \Gamma\left[ 1+\frac{K^2}{1-b}\left( r_i^\Delta+\frac{bH_b-1}{b} r_+^\Delta + \frac{r_i^{\Delta+\delta_0}}{b+1}r_+^{-\delta_0} + \cdots \right) \right], \label{h0hexp}
\end{eqnarray}
where $H_b$ is the $b^\text{th}$ harmonic number. Then $\eta/s$
can be obtained directly by the weaker horizon formula
(\ref{viscosityh+d+2}). As asserted in the main text, in the expansion (\ref{h0hexp}), the leading term is constant $\Gamma$ while scales $k^2$ and $e^{\phi_0}$ appear at the subleading term.
We find that when the axion
becomes irrelevant, the temperature dependence of $\eta/s$
 up to ${\cal O}(k^2)$ is more complicated than the case with (marginally)
relevant axion. The reason can be seen by rewriting the mass-like term as
\begin{equation}
K^2 r_+^\Delta\propto \left\{
\begin{array}{lll}
k^2 e^{\beta\phi_0}, & \text{for} & \Delta=0,\\
\left(\frac{k}{T^{1/z}}\right)^2 \left(\frac{e^{\phi_0}}{T^{\phi_1/z}}\right)^{\beta}, & \text{for} & \Delta\neq0,
\end{array} \right.
\end{equation}
where quantities $\{L_t,L_r,L_x\}$ have been absorbed in $\{T,e^{\phi_0},k\}$ by coordinate transformations. When axion is (marginally) relevant, namely, $\Delta=0$, the other
two scales $k$ and $e^{\phi_0}$ are combined into a scaleless
quantity $M^2$ and only enter the scaling dimension
(\ref{deltaT}). While, when axion is irrelevant, namely,
$\Delta\delta_0<0$, they enter $m(r)^2$ with a form coupling to
$T$ and lead to a complicated behavior of temperature dependence.
Nevertheless, since axion is irrelevant, $\eta/s$ is
finally expected to converge to a nonzero constant at extremal
low temperature \cite{Hartnoll:2016tri}. But, as seen from the
expansion (\ref{h0hexp}), the rate of convergence behaves
like $T^{\min(\frac{-\Delta}{z},\frac{\delta_0}{z})}$ that could happen to
be too slow to be observed numerically. The full expression (\ref{h0hirrelevant}) goes beyond the simple power law and
seems hard to obtain through scaling analysis.

In Eq.(\ref{h0hirrelevant}), there are two
parameters $\{\Gamma,r_i\}$ which should be given by fitting in the next subsection.

\subsection{Numerical calculation and fitting}
Now we conduct numerical calculation for neutral
background with positive specific heat and irrelevant axion. The
allowed parameter space is $z=1,\theta<0$ and $\Delta<0$
\cite{Gouteraux:2014hca}. Cases of $z\neq1$ can be constructed
with (marginally) relevant current \cite{Gouteraux:2014hca}. For
generality, we retain the freedom of $z$ in the following
discussion. Different from subsection
\ref{SubSectionUVCompletion}, the UV completive form of $V(\phi)$
is chosen as
\begin{equation}
V(\phi) =(d (d+1)-2 n_1) \left(1-\tanh ^2(\alpha  \phi )\right)+2 n_1 \cosh (\alpha  \phi ), \quad n_1=\frac{d \left(2 \alpha ^2+2 \alpha ^2 d+1\right)}{6 \alpha ^2}
\end{equation}
for the purpose that $V(\phi)$ approaches to $n_1e^{\alpha\phi}$
quickly even when $\phi$ is not too large. Other settings
in (\ref{action}) are $J(\phi)=e^{\beta\phi}, Z(\phi)=0$ and
$c=1$. The ansatz for the metric in numerical calculation
is similar to (\ref{ansatz})
\begin{equation}\label{ansatz2}
ds^2=\frac1{u^2}\left(-\frac{(1-u)U(u)}{S(u)}dt^2+\frac{du^2}{(1-u)U(u)}+ \sum_{i=1}^d dx_i^2 \right),\quad \phi=\phi(u),\quad \chi_i=kx_i.
\end{equation}
When $\phi$ is small, expansion $V(\phi)=d(d+1)+d\phi^2+\cdots$
gives boundary expansion $\phi=\tau u +\cdots + \nu u^{d} +
\cdots$. The boundary conditions are
$U(0)=1,\,S(0)=1,\,\phi'(0)=\tau$ at $u=0$ and regular conditions
at $u=1$. In this coordinate, temperature and entropy density are
$T=\frac{U(1)}{4\pi \sqrt{S(1)}}$ and $s=4\pi$. The dimensions of
$T$ and $k$ are chosen to be cancelled by unit $\tau$. Then
our numerical solutions are parameterized by two dimensionless
quantities $\{\frac T\tau,\frac k\tau\}$.

When lowering down $T/\lambda$, we fix $k/\lambda$.
Hyperscaling violation emerges near the horizon $u\to1$ at low
temperature. It can be directly observed by coordinate
transformation $u\to r^{1-\frac \theta d}/\tau,\, t\to
t/\tau,\,x_i\to x_i/\tau$ to the one used in (\ref{HVirrelevant}). Then we
know the location of horizon is $r_+=\tau^\frac
d{d-\theta}$ and the parameters accordingly transform as $T\to\tau
T,\,k\to \tau k$ and $s\to \tau^d s $. It means that the
temperature, lattices scale and entropy density in coordinate
(\ref{HVirrelevant}) are equal to the dimensionless ones in
(\ref{ansatz2}). Then the quantities
$\{L_r^2,L_t^2,e^{\phi_0}\}$  in (\ref{HVirrelevant}) can be
extracted from ansatz (\ref{ansatz2}) as
\begin{equation}\label{KLirr}
L_r^2=\frac{|d-\theta||\delta_0|}{d U(1)}\tau^\frac{-2\theta}{d-\theta},\quad
L_t^2=\frac{|d-\theta|U(1)}{d|\delta_0|S(1)}\tau^{\frac{2d(z-1)}{d-\theta}},\quad
e^{\phi_0}=e^{\phi(1)}\tau^\frac{-d\phi_1}{d-\theta}.
\end{equation}
Then $K^2$ can be calculated by using (\ref{h0irrelevant}).

\begin{figure}
  \centering
  \includegraphics[height=140pt]{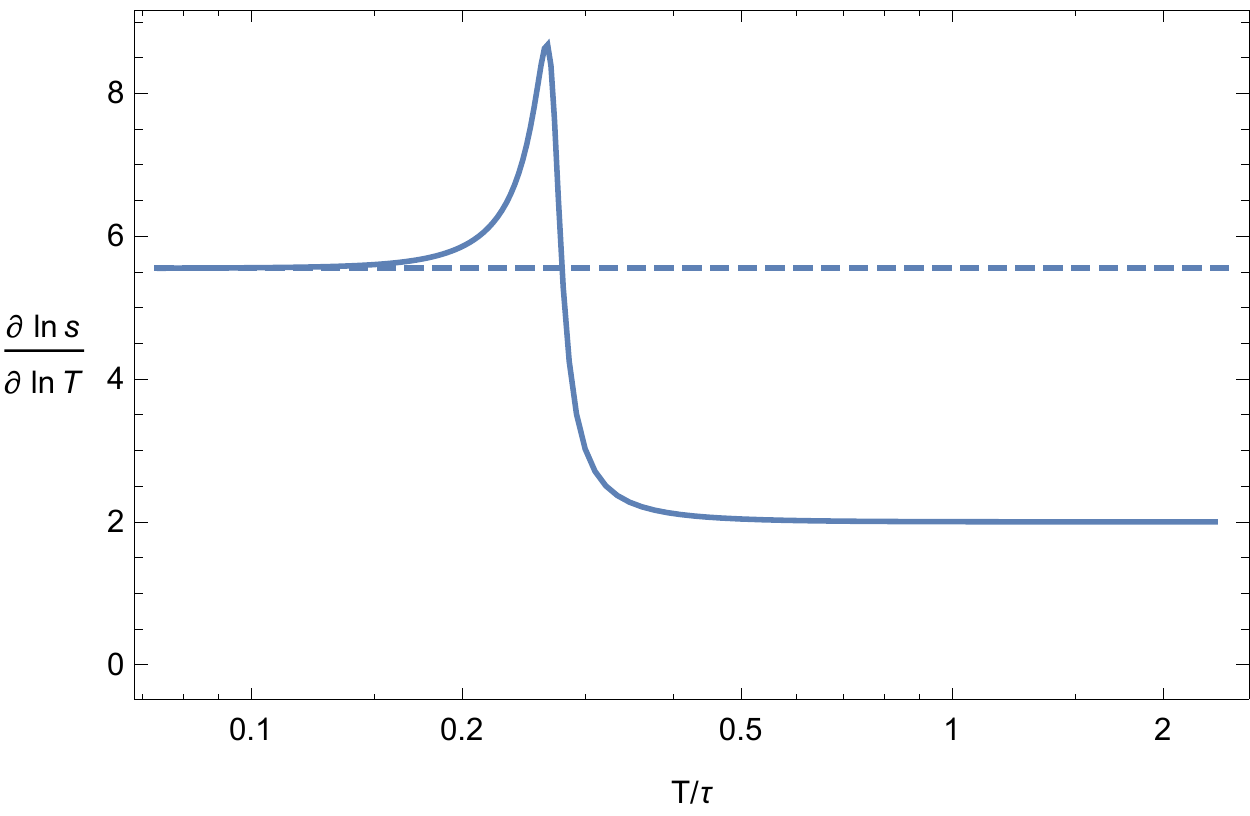}
  \includegraphics[height=140pt]{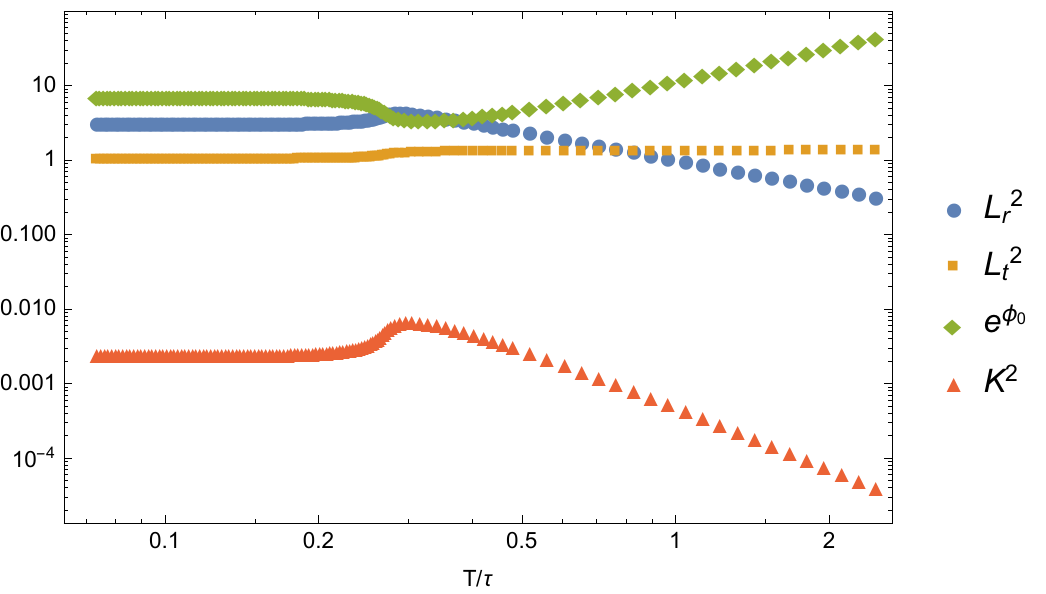}
  \caption{The exponent $\lambda$ of $s\sim T^\lambda$ as a function of $T/\tau$ is shown in the left plot, where solid line denotes numerical result and dashed line denotes analytical result. Quantities $\{L_r^2,L_t^2,e^{\phi_0},K^2\}$ as a function of $T/\tau$ is shown in the right plot. The parameters are $d=2,\,\alpha=0.8,\,\beta=-1$ and $k/\tau=0.5$, then $z=1,\,\theta=-3.56$ and $\Delta=-2.44$.
  }\label{sTirr}
\end{figure}

\begin{figure}
  \centering
  \includegraphics[height=140pt]{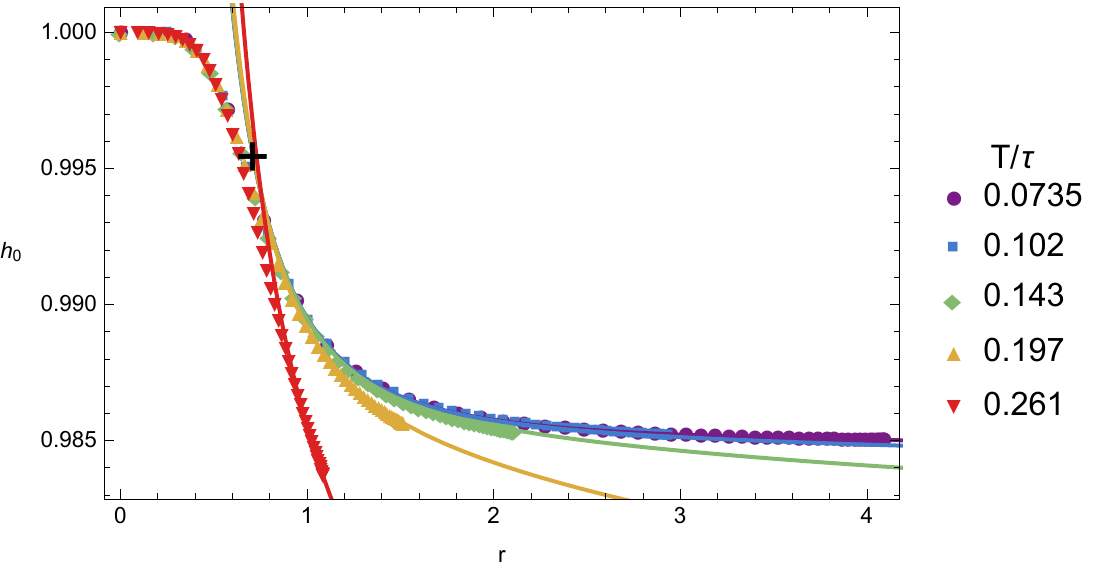}
  \includegraphics[height=140pt]{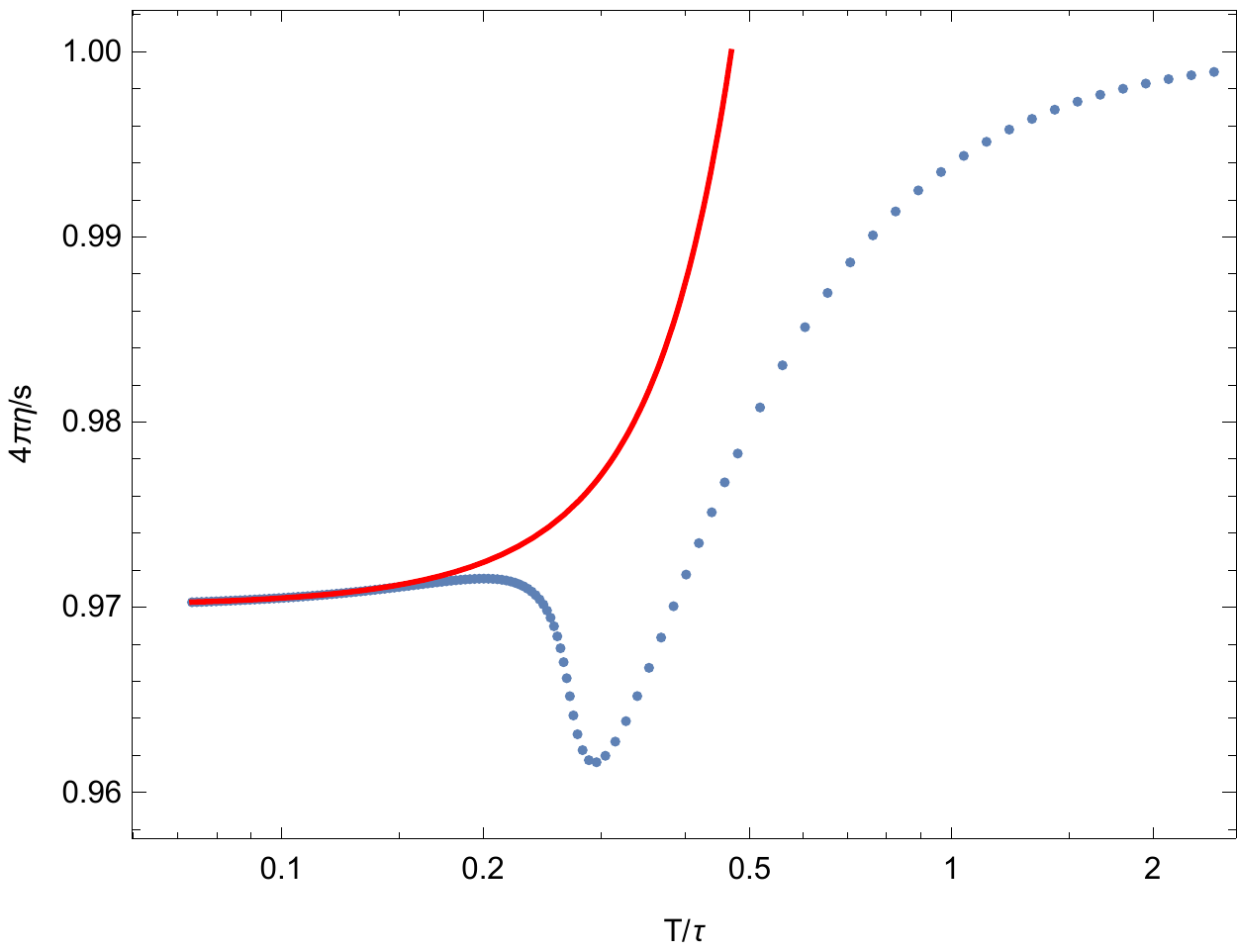}
  \caption{$h_0(r)$ at various temperatures are shown in the left plot, where the numerical solutions are denoted by points and the results from fitting (\ref{h0irrelevant}) are denoted by lines. $4\pi\eta/s$ as a function of $T/\tau$ is shown in the right plot, where blue points denote numerical result and red line denotes result from fitting (\ref{h0hirrelevant}). The fitting values of $\{\Gamma,r_i\}$ are marked by the black cross in the left plot.
  }\label{etaTirr}
\end{figure}

We demonstrate our numerical calculation in $d=2$. The
exponent $\lambda$ of $s\sim T^\lambda$ converges to
$\frac{d-\theta}{z}$ at low $T/\tau$, as shown in the left plot of
Figure \ref{sTirr}. Quantities
$\{L_r^2,L_t^2,e^{\phi_0},K^2\}$ go to constants at low $T/\tau$
as well, as shown in the right plot of Figure \ref{sTirr}.
Numerical solutions for $h_0(r)$ at different temperatures
$T/\tau$ are plotted in the left of Figure \ref{etaTirr}. As
expected, $h_0(r\lesssim1)$ in the UV region becomes
insensitive to $T/\tau$ when $T/\tau$ is small, while in the IR
region $h_0(r\gtrsim1)$ matches well with (\ref{h0irrelevant})
where constant $C$ is the fitting parameter. The temperature
behavior of $\eta/s$ is illustrated in the right plot of
Figure \ref{etaTirr}. From this figure we notice that
$\eta/s$ falls off quickly at first which is controlled by AdS
deformed by axion. It goes to the minimum at
$T/\tau\approx0.29$ then rises again because the axion
begins to be suppressed in the IR, which is also hinted by
the peak of $K^2$ at $T/\tau\approx0.29$ in Figure \ref{sTirr}.
When $T/\tau\approx0.2$, hyperscaling violation begins to emerge
and then $\eta/s$ begins to satisfy
(\ref{h0hirrelevant}). To fit the numerical data of $\eta/s$
by using (\ref{h0hirrelevant}), we fix the value of $K^2$ at lowest $T/\tau$ and select the fitting interval as $T/\tau<0.116$.
$\{\Gamma,r_i\}$ are the two fitting parameters, whose fitting values are marked in the left plot of Figure \ref{etaTirr}. The fitting curves match the data well when $T/\tau$ is low. We remark that the fitting values of $\{\Gamma,r_i\}$ are sensitive
to the fitting interval of $T/\tau$ but satisfy (\ref{connect})
pretty well at low $T/\tau$.

\end{appendix}

\end{document}